\RequirePackage[running]{lineno}

\documentclass[aps,prl,nofootinbib,twocolumn,superscriptaddress,letterpaper,amsmath,amssymb]{revtex4}

\pdfoutput=1

\usepackage{graphicx}  
\usepackage{dcolumn}   
\usepackage{bm}        
\usepackage{amssymb}   
\usepackage{feynmf}
\usepackage{slashed}
\usepackage{multirow}
\usepackage{hyperref}
\def\gsim{\lower0.5ex\hbox{$\:\buildrel >\over\sim\:$}}
\def\lsim{\lower0.5ex\hbox{$\:\buildrel <\over\sim\:$}}

\newcommand{\be}{\begin{equation}}
\newcommand{\ee}{\end{equation}}
\newcommand{\bea}{\begin{eqnarray}}
\newcommand{\eea}{\end{eqnarray}}

\newcommand{\nbox}{{\,\lower0.9pt\vbox{\hrule \hbox{\vrule height 0.2 cm
\hskip 0.2 cm \vrule height 0.2 cm}\hrule}\,}}

\def\sub#1{_{\lower.25ex\hbox{$\scriptstyle#1$}}}

\newskip\zatskip \zatskip=0pt plus0pt minus0pt
\def\matth{\mathsurround=0pt}
\def\lsim{\mathrel{\mathpalette\atversim<}}
\def\gsim{\mathrel{\mathpalette\atversim>}}
\def\sigv{\ifmmode \langle\sigma v\rangle\else $\langle\sigma v\rangle$\fi}
\newskip\zatskip \zatskip=0pt plus0pt minus0pt
\def\matth{\mathsurround=0pt}
\def\lsim{\mathrel{\mathpalette\atversim<}}
\def\gsim{\mathrel{\mathpalette\atversim>}}
\def\atversim#1#2{\lower0.7ex\vbox{\baselineskip\zatskip\lineskip\zatskip
  \lineskiplimit
  0pt\ialign{$\matth#1\hfil##\hfil$\crcr#2\crcr\sim\crcr}}}

\usepackage{graphicx,type1cm,eso-pic,color}



\begin{document}

\thispagestyle{empty}
\vspace*{-3.5cm}

\vspace{0.5in}

\title{ Observing Ultra-High Energy Cosmic Rays with Smartphones}

\begin{center}
\begin{abstract}
 We propose a novel approach for observing cosmic rays at ultra-high energy ($>10^{18}$~eV) by repurposing the existing network of smartphones as a ground detector array.  Extensive air showers generated by cosmic rays produce muons and high-energy photons,  which can be detected by the CMOS sensors of smartphone cameras. The small size and low efficiency of each sensor  is compensated by the large number of active phones. We show that if user adoption targets are met, such a network will have significant observing power at the highest energies.
\end{abstract}
\end{center}

\author{Daniel Whiteson}
\affiliation{ Corresponding author: daniel@uci.edu}
\affiliation{Department of Physics and Astronomy, University of
  California, Irvine, CA 92697}
\author{Michael Mulhearn}
\affiliation{Department of Physics, University of  California, Davis, CA }
\author{Chase Shimmin}
\affiliation{Department of Physics and Astronomy, University of
  California, Irvine, CA 92697}
\author{Kyle Cranmer}
\affiliation{Department of Physics, New York University, New York, NY}
\author{Kyle Brodie}
\affiliation{Department of Physics and Astronomy, University of
  California, Irvine, CA 92697}
\author{Dustin Burns}
\affiliation{Department of Physics, University of  California, Davis, CA }

\maketitle

\section*{Introduction}

The source of ultra-high energy cosmic rays (UHECR), those with energy above
$10^{18}$ eV, remains a puzzle even many decades after their discovery, as does the
mechanism behind their acceleration.  Their high energy leaves them less
susceptible to bending by magnetic fields between their source and the Earth,
making them excellent probes of the cosmic accelerators which produce
them~\cite{Abraham:2007si,Abraham:2007bb}.  But the mechanism and location of
this enormous acceleration is still not understood, despite many theoretical
conjectures~\cite{Bell:1978zc,Blandford:1987pw,Waxman:1995vg,Weiler:1997sh}.

When incident on the Earth's atmosphere, UHECRs produce
extensive air showers, which can be detected via the particle flux on the
ground, the flourescence in the air, or the radio and acoustic signatures.
A series of dedicated
detectors~\cite{Abbasi:2002ta,Takeda:1998ps,Abraham:2008ru} have detected cosmic
rays at successively higher energies, culminating in observation up to $3\cdot
10^{20}$ eV.  The flux of particles drops precipitously above $10^{18}$ GeV, due
to the suppression via interaction with the cosmic microwave
background~\cite{Greisen:1966jv,Zatsepin:1966jv}, making observation of these
particles challenging.

To accumulate a sufficient number of observed showers requires either a very long run or
a very large area.  Constructing and maintaining a new detector array with a
large effective area presents significant obstacles. Current arrays with large,
highly-efficient devices (Auger~\cite{Aab:2014ila}, Telescope Array~\cite{AbuZayyad:2012ru}, AGASA~\cite{Hayashida:1998qb}) cannot grow dramatically larger without
becoming much more expensive. Distributed detector arrays with small,
cheaper devices ({\it e.g.} ERGO~\cite{ergo}) have the potential to grow very large, but
have not achieved the size and density required to probe air showers,
potentially due to the organizational obstacles of production, distribution and
maintenance of their custom-built devices.

It has been previously shown that smartphones can detect ionizing radiation~\cite{Cogliati:2014uua,2002SPIE.4669..172S,decoaps}. In this paper, we demonstrate that a dense network of such devices has power sufficient to detect air showers from the highest energy cosmic rays.  We measure the particle-detection efficiency of several popular smartphone models, which is necessary for the reconstruction of the energy and direction of the particle initiating the shower. With sufficient user adoption, such a distributed network of devices can observe UHECRs at rates comparable or exceeding conventional cosmic ray observatories.  Finally, we describe the operating principles, technical design and expected sensitivity of the {\sc CRAYFIS} (Cosmic RAYs Found In Smartphones) detector array. Preliminary applications for Android and iOS platforms are available for testing~\cite{website}.

\section*{Detection}

\begin{figure}[htbp]
\includegraphics[width=3.2in]{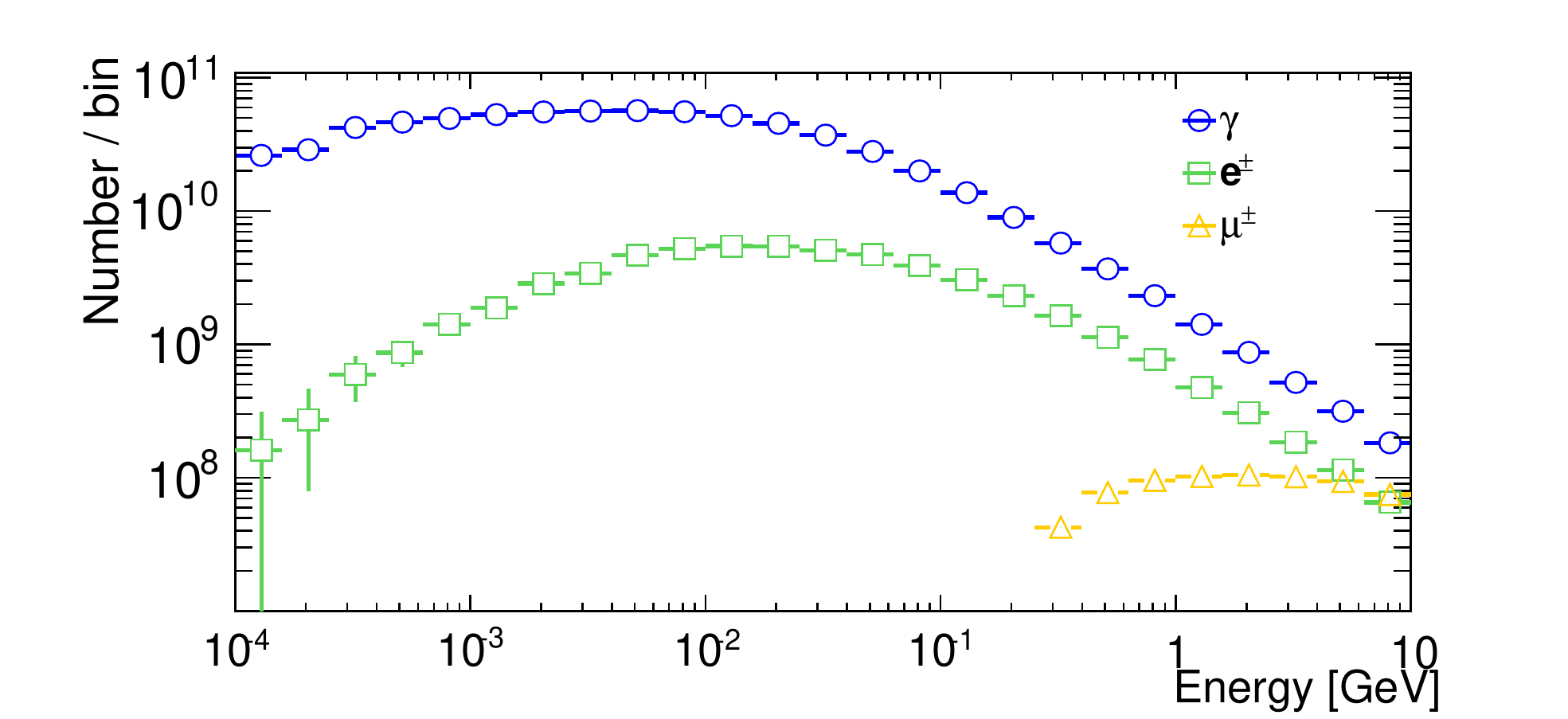}
\includegraphics[width=3.2in]{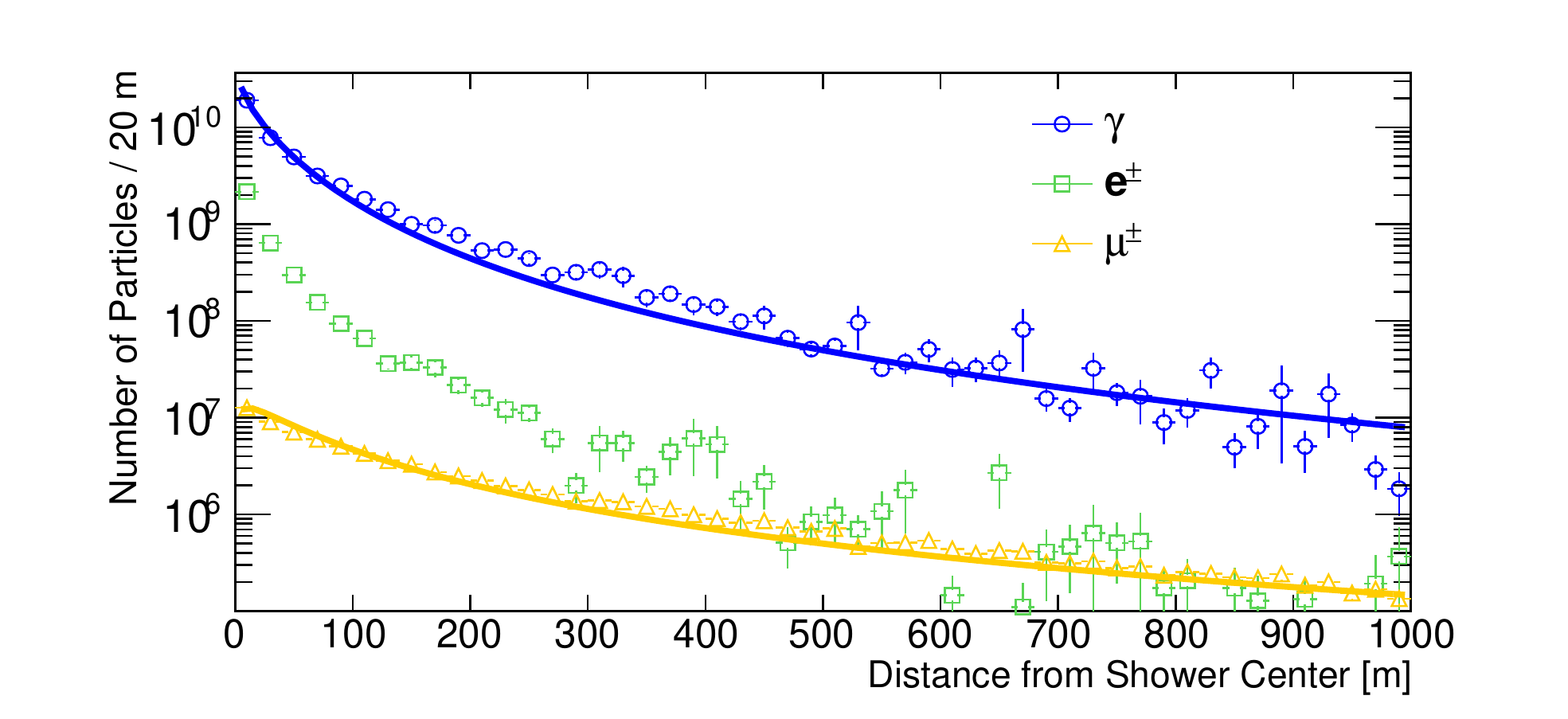}
\caption{Energy spectrum (a) and distance from shower axis (b) of photons, electrons, and muons at ground level for simulated air showers initiated by protons with energy $10^{19}-10^{20}$ eV. Also shown (b) is a parametric fit to Eq~\ref{eq:ldf}.}
\label{fig:eas}
\end{figure}

Air showers induced by cosmic rays contain an enormous number of particles. Figure~\ref{fig:eas} shows the energy spectrum, and spatial distribution at sea level of photons, electrons, and muons in showers as simulated by the {\sc Corsika}~\cite{Heck:1998vt} program with the {\sc QGS-II}~\cite{Ostapchenko:2010vb} model of hadronic interactions.

We focus our attention on photons, which have high densities in the shower, and
muons, which have excellent penetrating power and high detection efficiency.
Electrons are also numerous and have high efficiency, but may be blocked by buildings, phone cases or camera lenses.  Heavier hadronic particles are much less common. 

The sensitive element in a smartphone is the camera, a CMOS device in which silicon photodiode pixels produce electron-hole pairs when struck by visible photons. While these devices are designed to have reasonable quantum efficiency for visible light, the same principle allows the sensor to detect higher-energy photons~\cite{Cogliati:2014uua} as well as minimally-ionizing particles such as muons~\cite{Matis:2002jv, Kleinf2002}.  

Our own {\sc Geant}-based simulation~\cite{Agostinelli:2002hh} of muons, protons, electrons and photons incident on a simple block of Silicon confirms that there is significant interaction and energy deposition for incident particles in the energy range expected in an air shower. The modeling of the microphysical processes involved have been validated extensively in the context of silicon-strip detectors for particle physics experiments, though the specifics of the doping and electric-field configuration needed for readout differ substantially between collider silicon-based detectors and CMOS devices and have not been included in this initial study. 

An application running on the smartphone has access to an array of {\it pixel response} values, commonly with eight-bit precision.  Though many stages of processing occur between the direct measure of the deposited energy by the CMOS sensor and the delivery of pixel response values, we assume that the former is a reasonable proxy for the latter.

\section*{Software}

With the camera as the detector element, the phone processor runs an application
which functions as the trigger and data acquisition system.  To obtain the largest possible integrated exposure time, the first-level
trigger captures video frames at 15-30 Hz, depending on the frame-processing speed of the device. Frames which contain any above-threshold pixels are stored and passed to the second stage which examines the stored frames, saving
only the pixels above a second, lower threshold. 
All qualifying pixels, typically a few per frame, are stored as a sparse array in a buffer on the phone, along with their arrival time and the geolocation of the phone.
When a wi-fi connection is available, the collected pixels are uploaded to a central server for offline shower reconstruction; most events are between $50-200$ bytes of data.
The acquired event rate may be tuned by setting the thresholds to eliminate spurious background events; typical rates are around $0.2$ Hz.

The application is designed to run when the phone is not in active use and a power source is available.  No additional light shielding of the camera, such as tape, is required, other than placing the phone face-up (camera-down) on a table. In a few devices tested, performance at night-time without shielding was equivalent to tests done with shielding; this performance may vary with model type. Real performance will vary according to the device camera geometry as well as the ambient conditions; per-device calibration will be important in establishing backgrounds levels.  In
this way, no active participation is required once the application is installed
and its operation should be fairly unobtrusive, which is critical to achieving
wide participation in the smartphone community.  To address user privacy concerns, no frames will be stored or uploaded if the average pixel response value over the frame exceeds a threshold, such that full images cannot be reconstructed offline.

Offline, we perform hot-pixel removal. Individual pixels that fire at a rate much higher than the average are removed;
these are caused either by light leakage, typically near the edge of the frame, or by poorly-performing or noisy pixels.

\section*{ Photon Reconstruction and Efficiency}

Detection of particles in smartphones has been performed previously~\cite{Cogliati:2014uua}, but application of such measurements to the observation of extensive air showers from UHECRs has not yet been explored.  A critical step is understanding the product of active area and the efficiency $A\epsilon$ of each device for the particles species in an air shower.  The number of events $N_{\rm cand}$ that pass the trigger threshold  determines the efficiency $\epsilon = N_{\rm cand} / N_{\rm incident}$ of the device.   Measurements of the efficiency are presented below, and details of $A$ are typically available from manufacturers.

\begin{figure}
\includegraphics[width=3in]{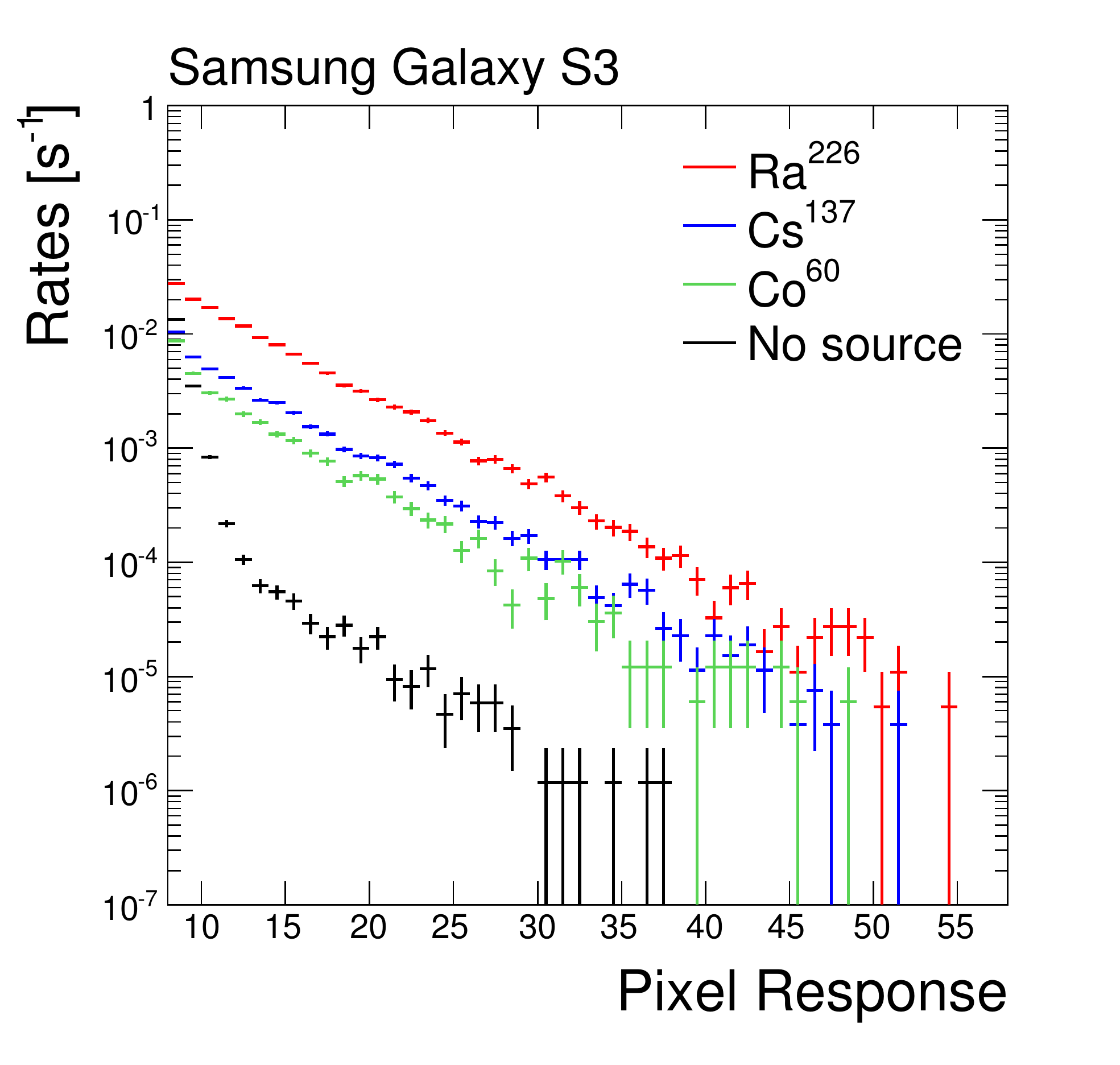}
\caption{ Distribution of observed pixel response values in a Samsung Galaxy S3 phone when exposed to sources which emit photons between 30-1200 keV, and without any source.  The differences in rates are due to the different activity of the sources. The data with no source shows a falling noise distribution and a tail attributed to cosmic muons. Other phone models show qualitatively similar behavior.}
\label{fig:pixval}
\end{figure}

\begin{figure}
\includegraphics[width=2.5in]{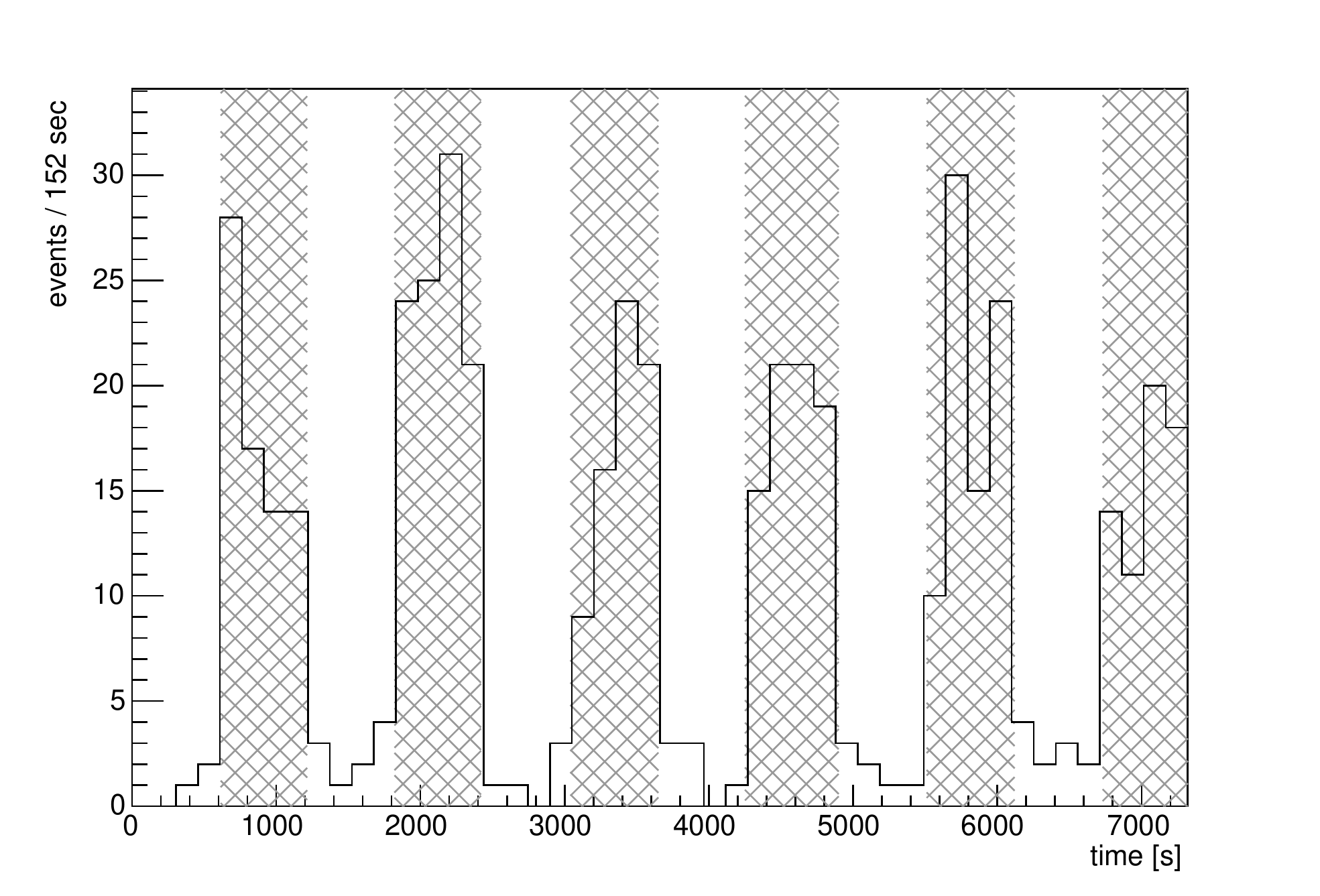}
\caption{Number of pixels with value above trigger threshold versus time in a Samsung Galaxy S3 phone. In the periods indicated by hatching, a $^{226}$Ra source has been placed near the phone.}
\label{fig:exg}
\end{figure}

\begin{figure}
\includegraphics[width=1.5in]{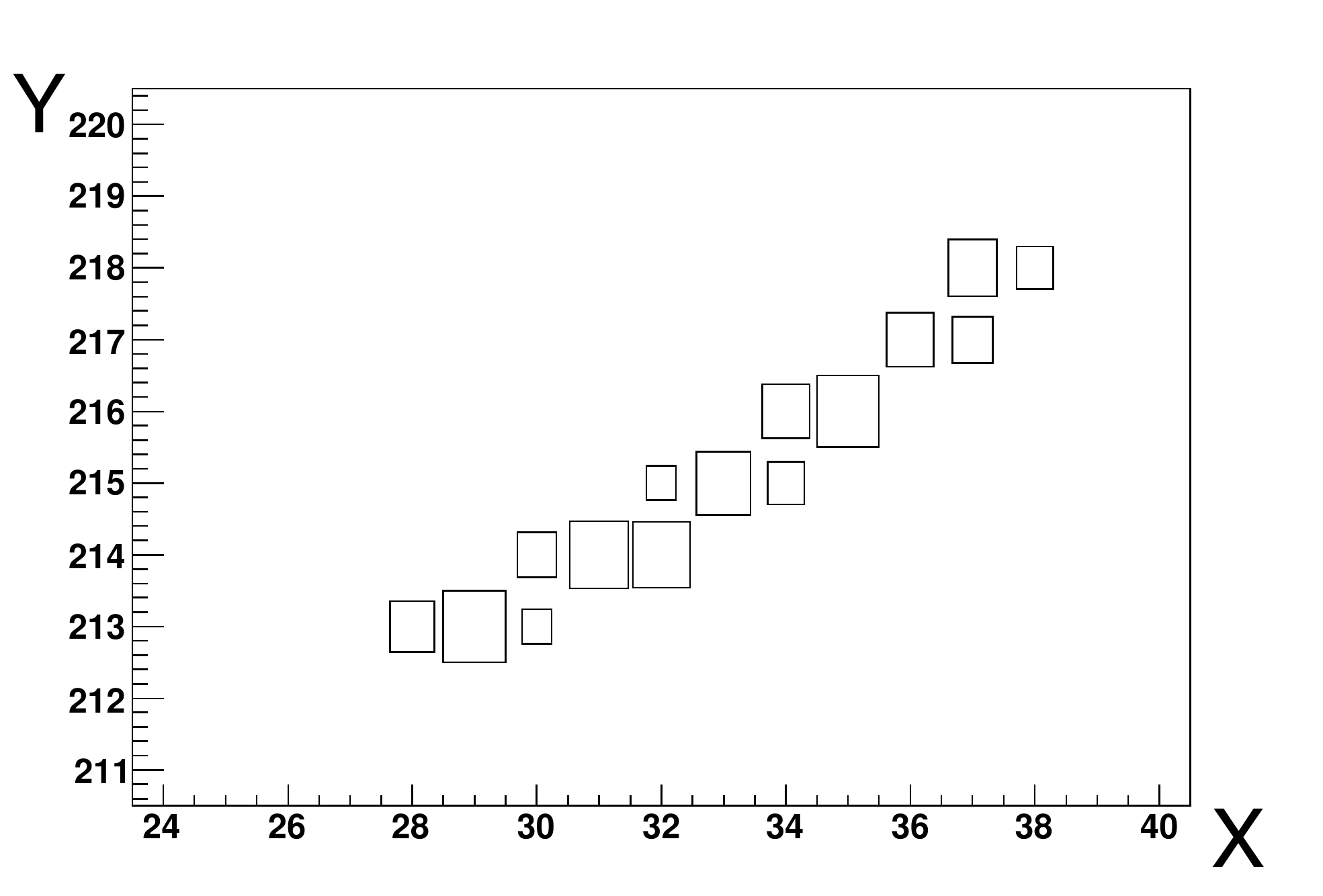}
\includegraphics[width=1.5in]{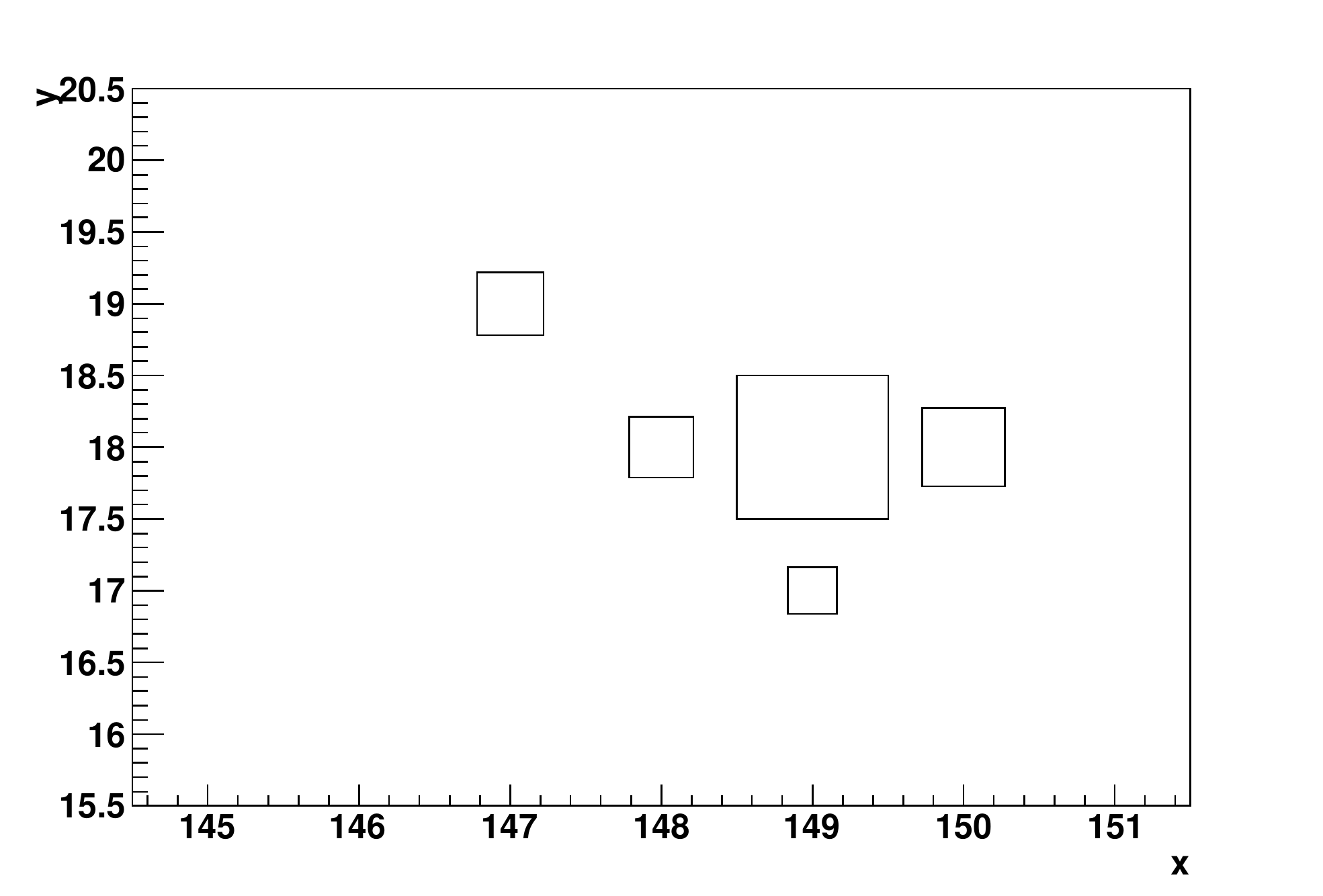}
\includegraphics[width=1.5in]{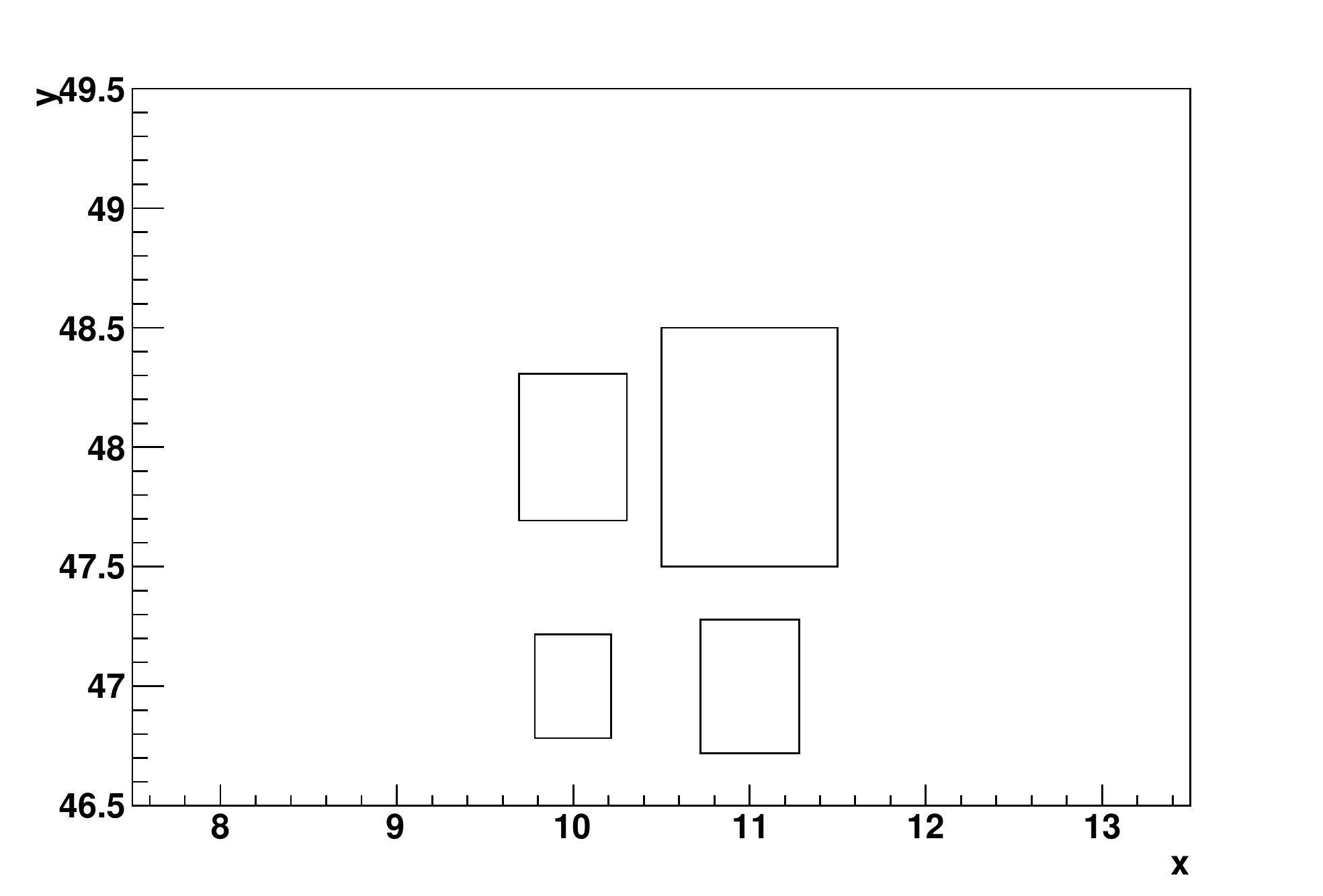}
\includegraphics[width=1.5in]{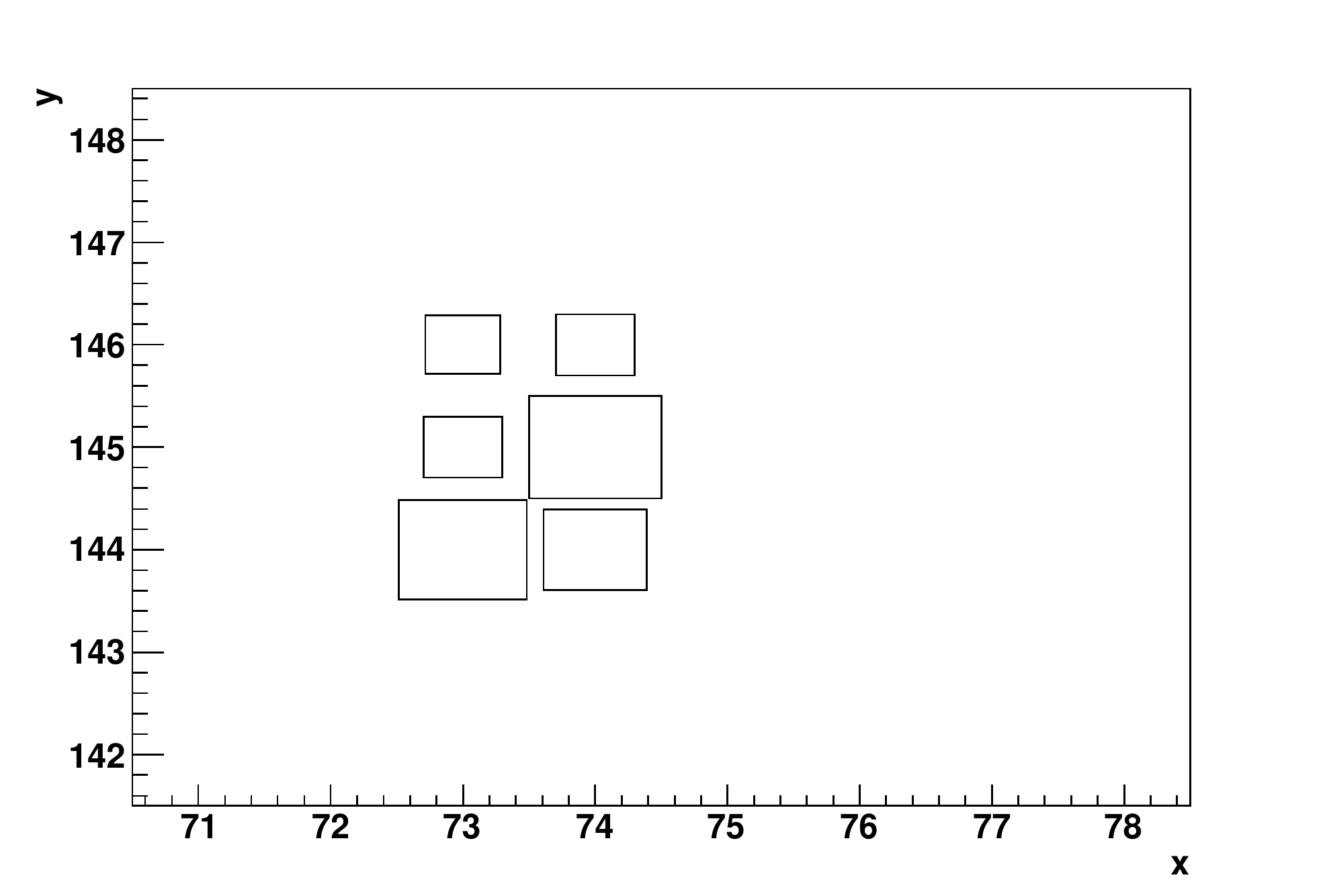}
\caption{Activated pixels above threshold in a Samsung Galaxy SIII phone, during exposure to $^{60}$Co. Box size is proportional to pixel response values}
\label{fig:clusterg}
\end{figure}

The response of several popular phone models to photons was measured in the lab
using gamma rays from the radioactive decays of Ra$^{226}$ ($E_{\gamma}=30-600$
keV), Co$^{60}$ ($E_{\gamma}=1.2-1.3$ MeV) and Cs$^{137}$ ($E_{\gamma} \le 700$
keV).  As a representative example, the measured pixel response of a Samsung Galaxy S3 is shown in
Fig.~\ref{fig:pixval}; similar spectra are seen in other Android models as well as iPhones.  In the presence of radioactive sources, the camera
detects pixels with a large charge deposition at a rate that is proportional to
the activity of our sources.  When no source is nearby, the distribution of pixel response values  presents a steeply falling distribution, with a long tail that we attribute to
cosmic ray muons (see discussion below). To confirm the  sensitivity of the phones to photons, we periodically place a radioactive source near a phone and remove it;  Fig.~\ref{fig:exg} shows that the number of pixels with a value
above a trigger threshold is highly correlated with the presence of a
radioactive source.
In addition to leaving isolated pixels with large pixel response values, some high energy photons leave several bright pixels in clusters or tracks; see Fig.~\ref{fig:clusterg}. These can be attributed to compton-scattered and pair-produced electrons.

For a radioactive source with activity $R$ a distance $d$ from the sensor, we can measure $A\epsilon$ by counting the number of events observed $N_{\rm obs}$ over a period $\Delta t$:
\[ A\epsilon  = 4\pi \, d^2 \frac{N_{\rm obs}}{R \Delta t}. \]

The activity of each radioactive source was determined using a high-precision photon counter at the UC Irvine test reactor.
The distance from the camera to the source was kept constant using a wax assembly, allowing us to measure $A \epsilon $ to within a few percent.
We found only minor variation in $A\epsilon$ for the different photon sources listed above.
Between the different phone models tested, we measured a range of $A\epsilon$ of $2.5\times 10^{-9} - 2.5\times 10^{-8}$ m$^2$, with consistent values of $A\epsilon$ between phones of the same model.
We therefore consider a conservative range of average photon sensitivity of $A\epsilon = (1-5)\times 10^{-9}$ m$^2$ for projections. Note that improved background suppression at the trigger level could yield increased photon candidate efficiency.

\section*{ Muon Reconstruction and Efficiency}

\begin{figure}
\includegraphics[width=2.5in]{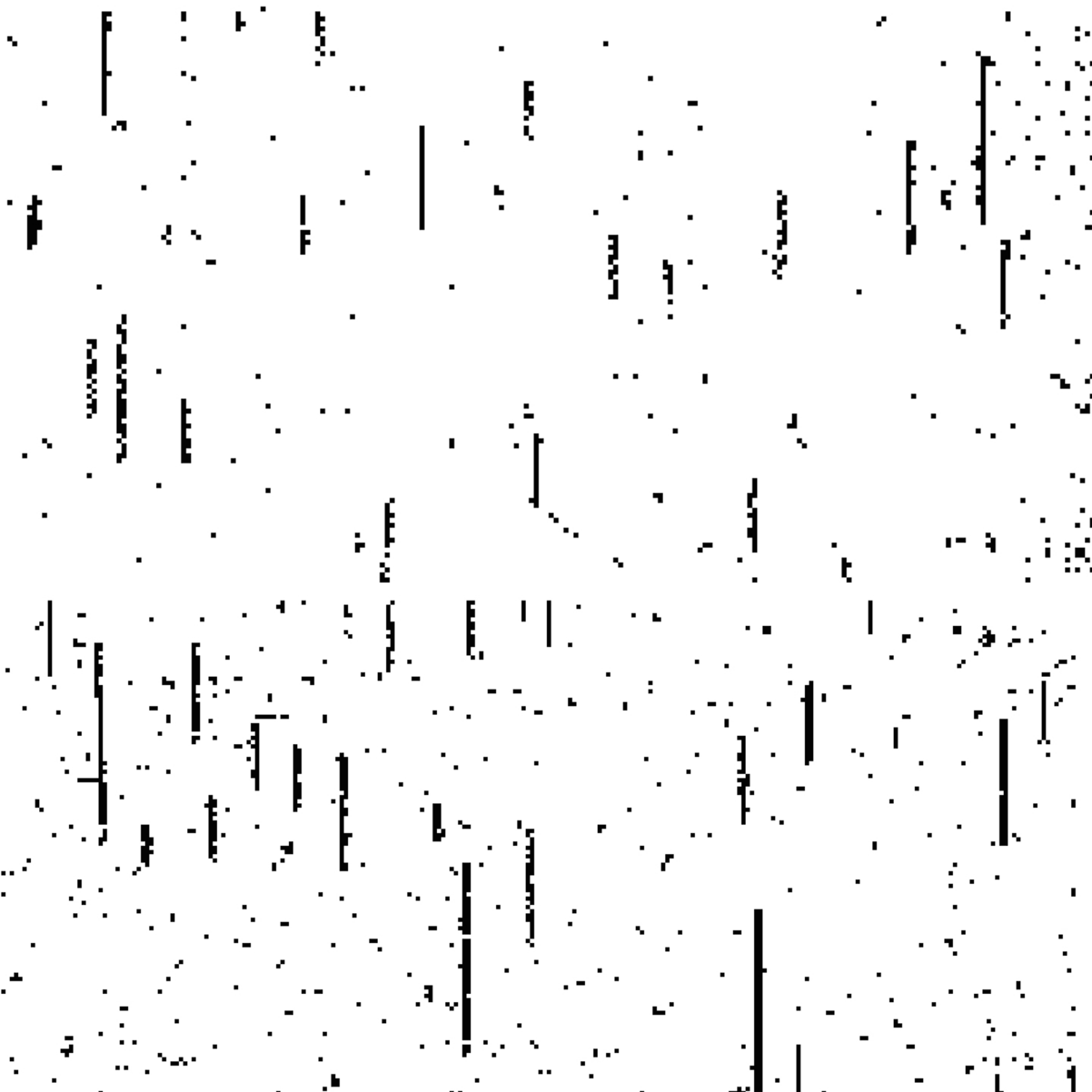}
\caption{ Composite image of activated pixels in data collected from phones exposed to a muon beam. The phones were arranged such that the muon beam was incident on the side of the sensor, giving visible tracks where muons pass through several pixels.}
\label{fig:exm}
\end{figure}

Muon efficiencies for each device are calculated by comparing the rate
of candidate muons to the expected rate from cosmic ray muons.  In
surface experiments, at least 5 cm of lead shielding was used to
suppress contributions from ambient radioactivitiy.  Runs at higher
altitude during commercial airline flights display an increase in
observed charged particle candidates, consistent with expectation.
However, the uncertainty in both the actual local muon flux and the
fraction of candidate muons which are due to other sources (such as
electronic noise or background radioactivity) lead to  large
uncertainty on the muon efficiency. For this reason we
consider two benchmark muon $A\epsilon$ scenarios which reflect the uncertainty in the muon efficiency and variation in sensor sizes.  Composite images from
phones exposed to a muon beam at CERN in Geneva, Switzerland are
shown in Fig.~\ref{fig:exm}.  The beam was incident on the side of the
phone, and the image has clear muon tracks from that direction; the nearly unbroken nature of these tracks implies a fairly high efficiency.
Studies using a muon telescope to tag muon candidates and
directly measure the per-pixel efficiency to muons are currently underway.

\section*{Shower Reconstruction}

In the presence of an air shower, the local density of particles can be written
as a vector ${\bm \rho}(x,y)$, where each component refers to a particular
species of particle (muon or photon).  A phone, with active detector element
area $A$, and particle species identification efficiency vector ${\bm \epsilon}
= (\epsilon_\mu,\epsilon_\gamma)$, will reconstruct a mean number of candidates
  $\lambda = \eta + A {\bm \epsilon} \cdot {\bm \rho}(x,y)$, where $\eta$ is the
  expected number of uncorrelated hits from background sources  due to electronic noise, uncorrelated muons and ambient radioactivity in a coincidence window.

\begin{figure}[hbt!]
\includegraphics[width=3.0in]{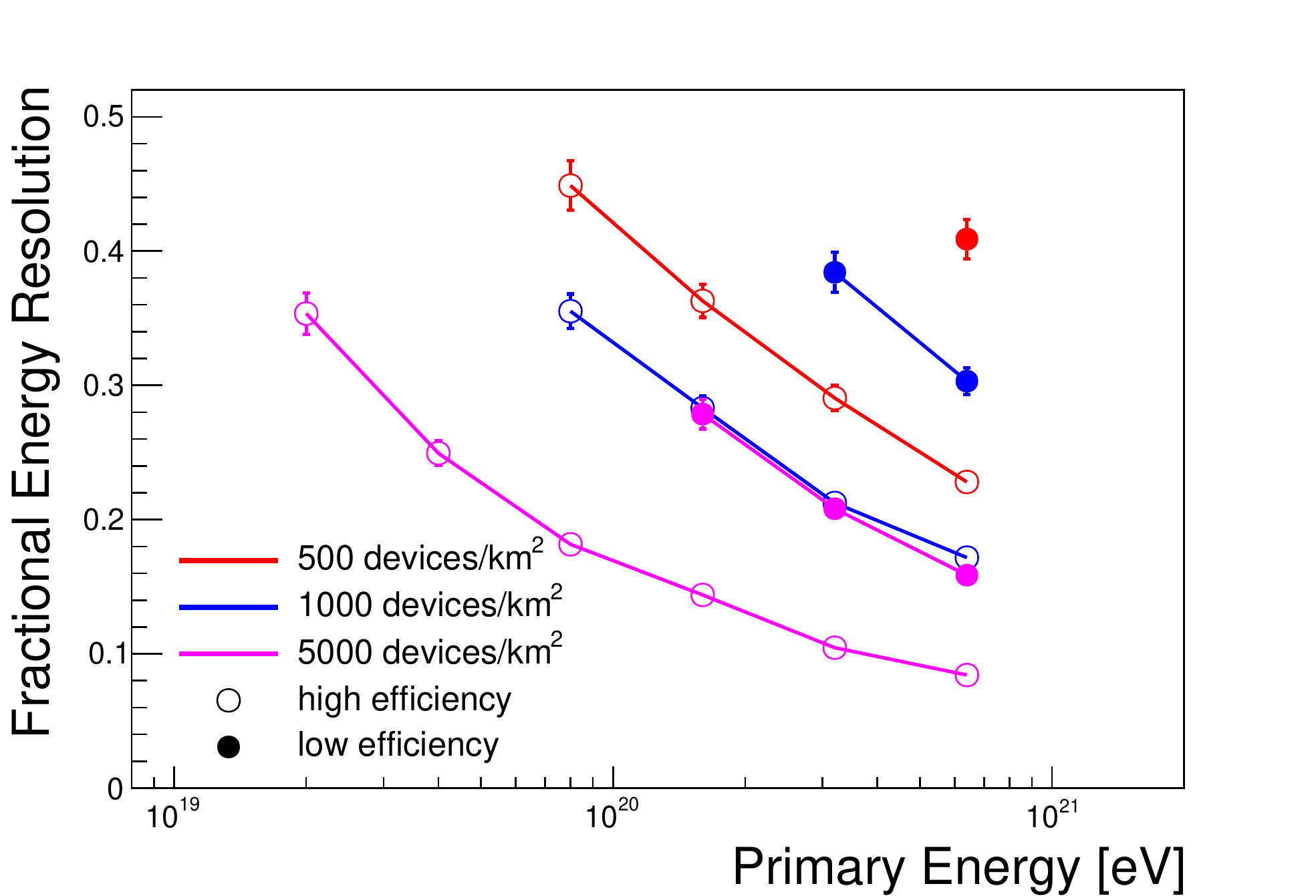} \\
\includegraphics[width=3.0in]{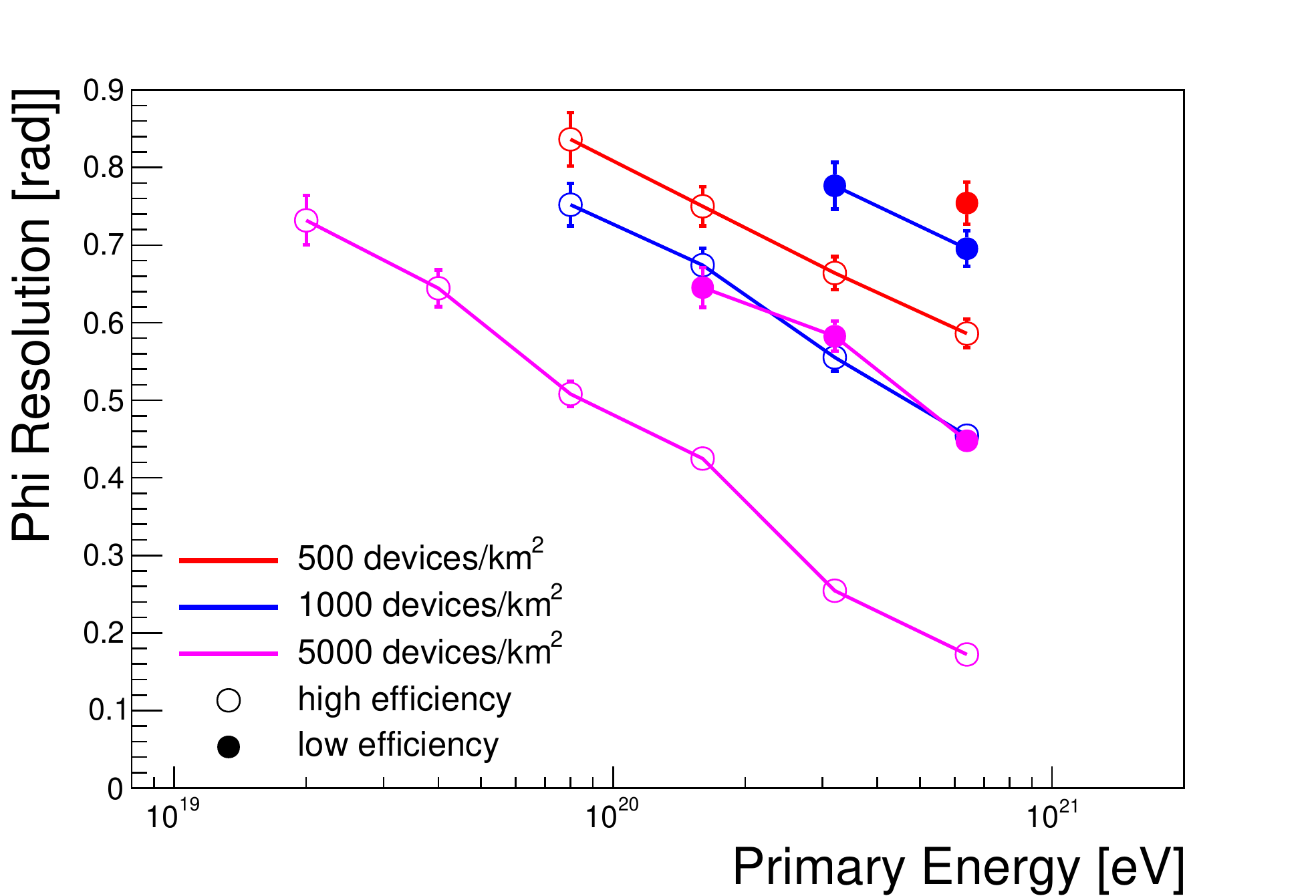} \\
\includegraphics[width=3.0in]{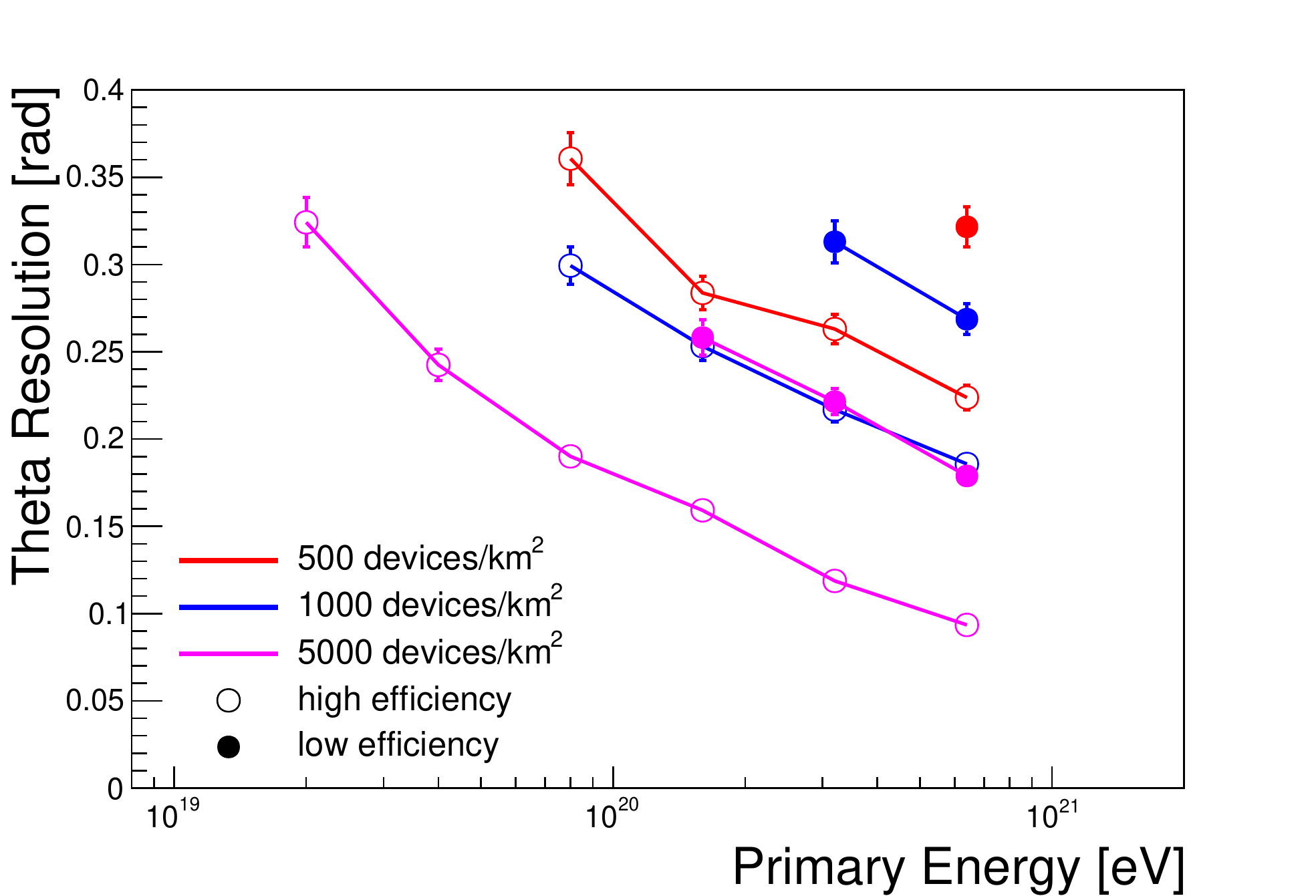} \\

\caption{ Fractional energy resolution (top), and $\phi$ (middle) and
  $\theta$ (bottom) resolution in radians, for simulated events.  In
  the high-efficiency scenario $A\epsilon  = 5 \times
  10^{-5}~\rm{m^2}$ for muons and $5 \times 10^{-9}~\rm{m^2}$ for
  gamma rays.  In the low-efficiency scenario $A\epsilon  = 1
  \times 10^{-5}~\rm{m^2}$ for muons and $1 \times 10^{-9}~\rm{m^2}$
  for gamma rays.  Also shown are three different device densities.}
\label{fig:fitperf}
\end{figure}

The probability that a phone will register no candidates is then given by the Poisson distribution
\begin{displaymath}
 P_{0}(x,y) = e^{-A {\bm \epsilon} \cdot {\bm \rho(x,y)}-\eta},
\end{displaymath}
and the probability that the phone will register one or more candidates is 
\begin{displaymath}
 P_{1}(x,y) = 1- e^{-A {\bm \epsilon} \cdot {\bm \rho}(x,y)-\eta}.
\end{displaymath}

\noindent
If the quantity $A\epsilon$ is known for each phone and particle
species type, measuring the distribution of phones with and without
candidates constrains the local shower density ${\rho}_i (x,y)$, of each
particle species $i$. The density is directly proportional to the incident
particle energy with a distribution in $x$ and $y$ sensitive to the
incident particle direction.   We use a parameterized model for $\rho$~\cite{grieder}
\begin{eqnarray}
\rho(N_i,r,s) = \frac{N_i}{2\pi r^2_M}\left (\frac{r}{r_M}\right)^{(s-2)} \left (1+\frac{r}{r_M}\right)^{(s-4.5)} \nonumber \\
 \times \left ( \frac{\Gamma(4.5-s)}{\Gamma(s)\Gamma(4.5-2s)}\right ) [\textrm{m}^{-2}] 
\label{eq:ldf}
\end{eqnarray}
where $r$ is the distance of a detector element to the vector of the original
incident particle,
$r_M$ is the Moliere radius in air,
$s$ is the shower age ($s=1$ being the shower maximum) and $N_i$ is the number of
particles of species $i$ in the shower.  This parameterization has been validated in realistic
simulations from {\sc Corsika}, see Fig~\ref{fig:eas}. This approach neglects some sources of systematic uncertainty, such as the hadronic interaction model, fluctuations of the shower shape, dependence on the atmospheric conditions, and dependence of $\rho$ on the initial particle species. Therefore, the resolutions quoted below should be seen as an optimistic lower bound; performance of a real network is likely to be significantly worse.

We use an unbinned likelihood to extract incident particle energy and direction:

\begin{displaymath}
L(N,\theta,\phi) = \prod_{i} P_{\textrm{0}}(x_i, y_i) \prod_{j} P_{1}(x_j, y_j)
\end{displaymath}

\noindent where the $i$ index runs across phones that did not reconstruct a
candidate and the $j$ index runs across phones that did reconstruct a
candidate.  The symmetric use of phones with and without candidates naturally
handles the non-uniform distribution of participating phones.  In areas of high
particle density, the possibility exists of multiple hits on a single phone, allowing for
additional power in determining the shower density. We leave this for later
refinements.  

Expected performance in simulated events drawn from Eq.~\ref{eq:ldf} is shown in
Fig.~\ref{fig:fitperf} for various scenarios in $A\epsilon$ and phone density.
The resolution improves with higher shower energy due to higher statistics from
an increased particle density.  Lower values of $A\epsilon$ can be compensated by
higher phone densities, as shown by the overlapping curves.

\begin{figure}
\includegraphics[width=3.0in]{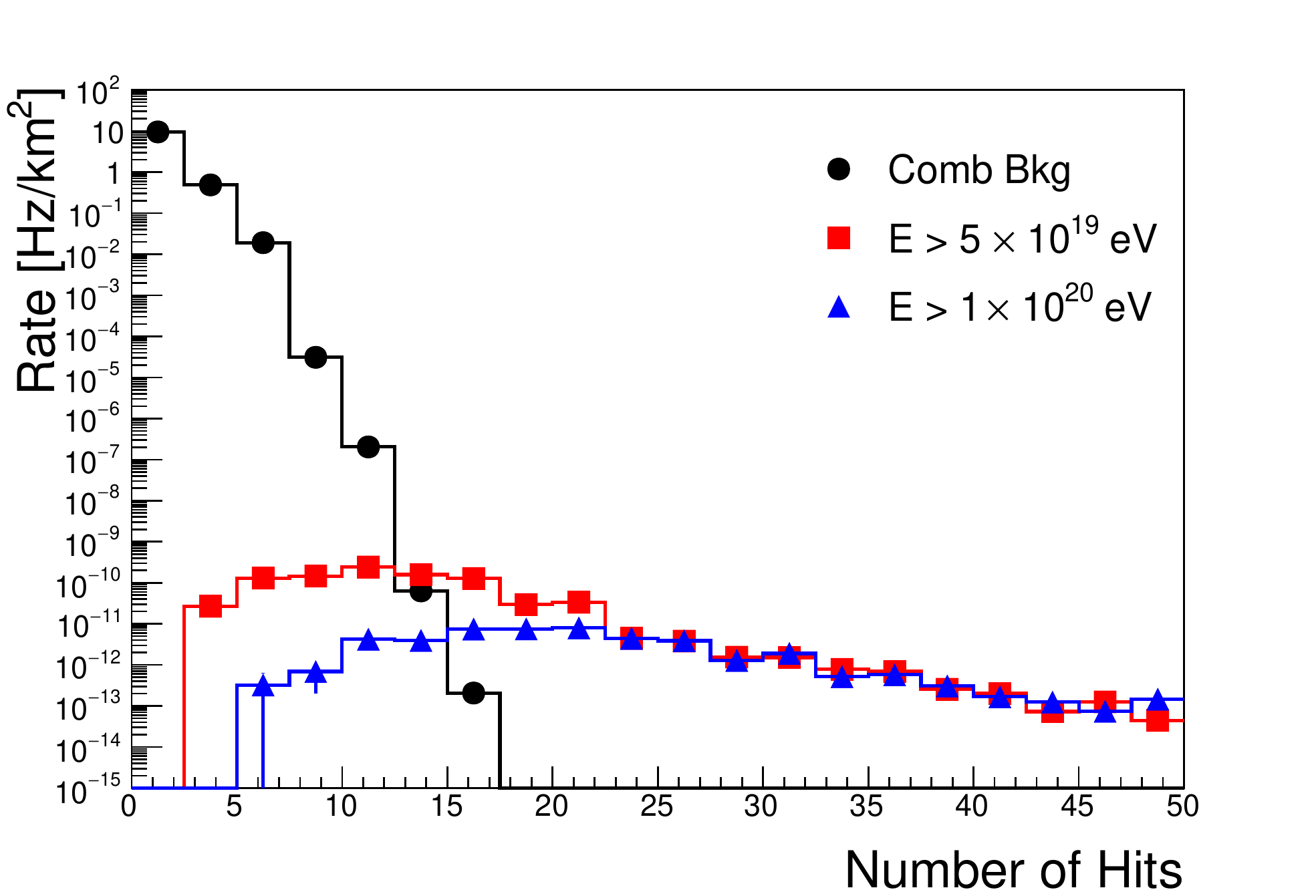}\\
\includegraphics[width=3.0in]{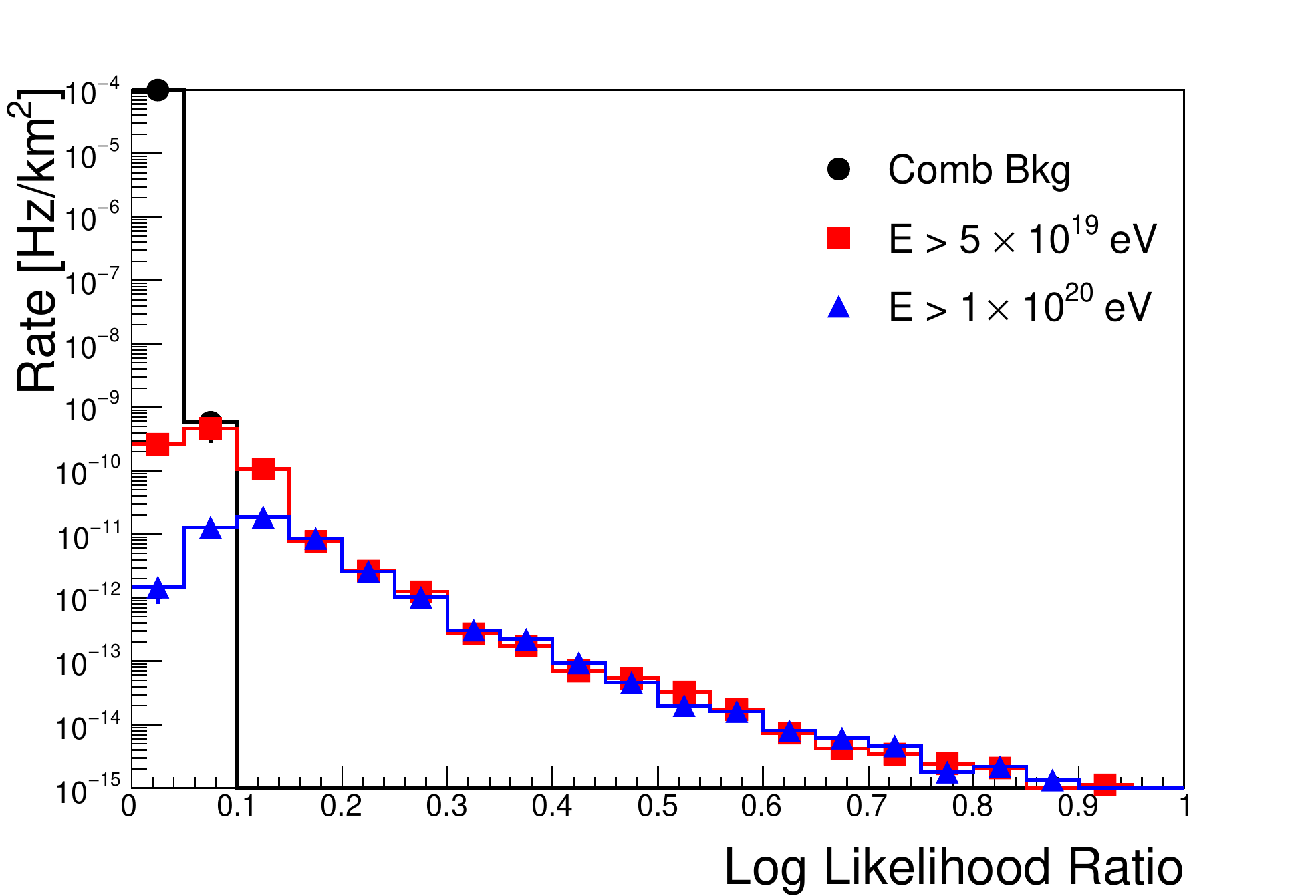}\\
\includegraphics[width=3.0in]{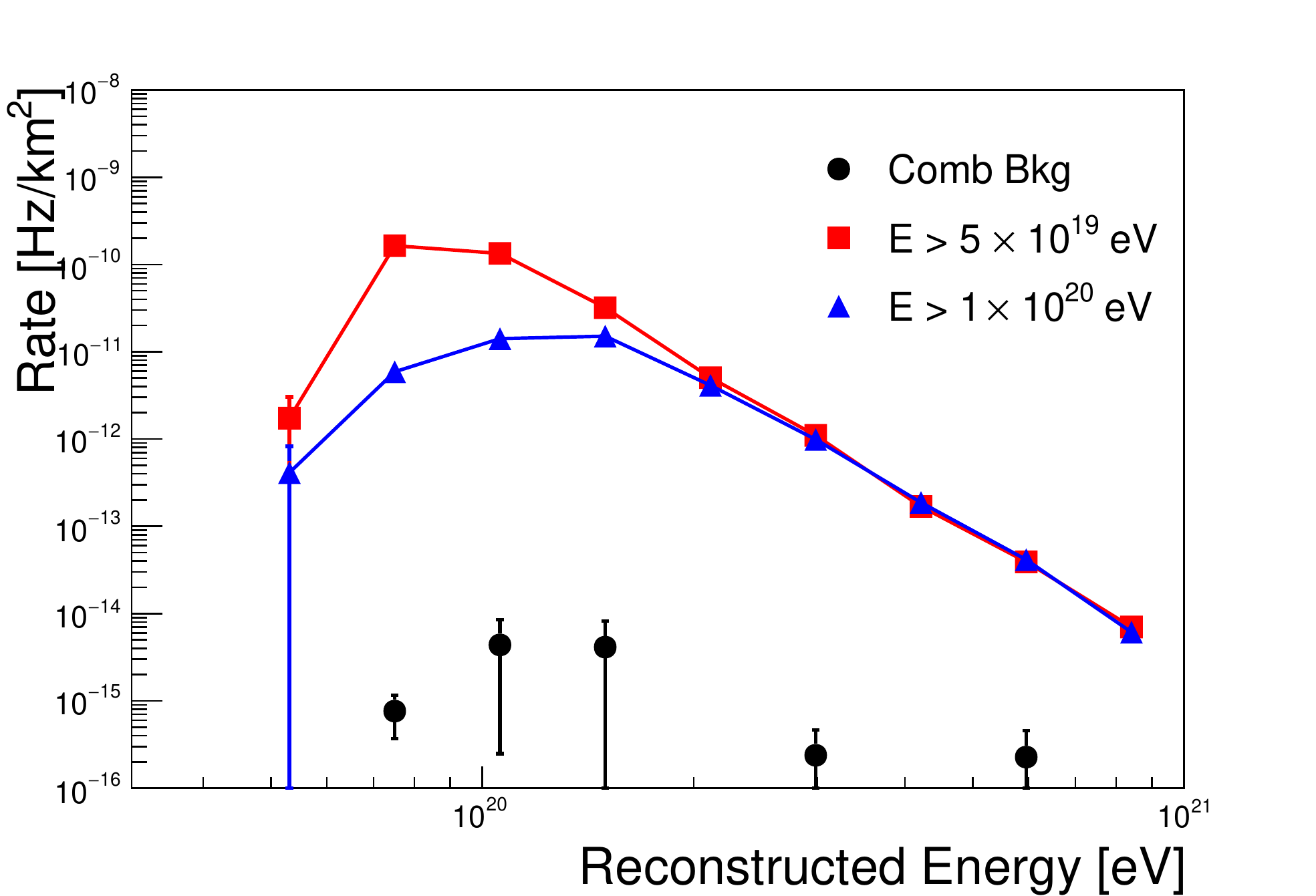}\\
\caption{ \label{fig:noise} For a $1~{\rm km}^2$ grid, the expected
  observation rate of (top) total number of phones registering a hit,
  (middle) log likelihood ratio of the best fit shower to pure
  combinatorial background, and (bottom) reconstructed primary energy, for
  irreducible combinatorial background only, primary showers with
  energy greater than $5\times 10^{19}$ eV, and primary shower with energy greater
  than $1\times 10^{20}$ eV. These plots assume the optimistic muon efficiency benchmark 
  ($A \epsilon = 5\times 10^{-5}~$m$^2$) and a timing resolution of 100~ms.}
\end{figure}

The background, due to electronic noise, uncorrelated muons and
ambient radioactivity, is not expected to be correlated among phones.
Assuming the background rate is dominated by cosmic ray muons with
flux of $1\, \text{cm}^{-2}\, \text{min}^{-1}$, the expected number of
uncorrelated background hits during a 100~ms coincidence window in the
high-efficiency scenario is $\eta=0.0008$.  Fig.~\ref{fig:noise}
compares the rate of several experimental observables for pure
combinatorial background and simulated showers with primary energy
$E>5\times 10^{19}$ eV and $E>1\times 10^{20}$ eV. The rate of
high-energy showers is many orders of magnitude lower than the rate at
which candidate events are acquired, so extremely rare coincidences can
produce a relatively large number of phone hits at rates comparable to
high-energy showers.  However, the log likelihood ratio (LLR) between
the best fit shower and pure combinatorial background is a powerful
discriminant. The rate of combinatorial background events depends on
the timing resolution of the devices.  Similar studies have been
performed for larger coincidence windows with qualitatively similar
results, though the rate of such background events rises with larger
windows.

\section*{Expected Observational Power}

\begin{figure}[t!]
\includegraphics[width=3in]{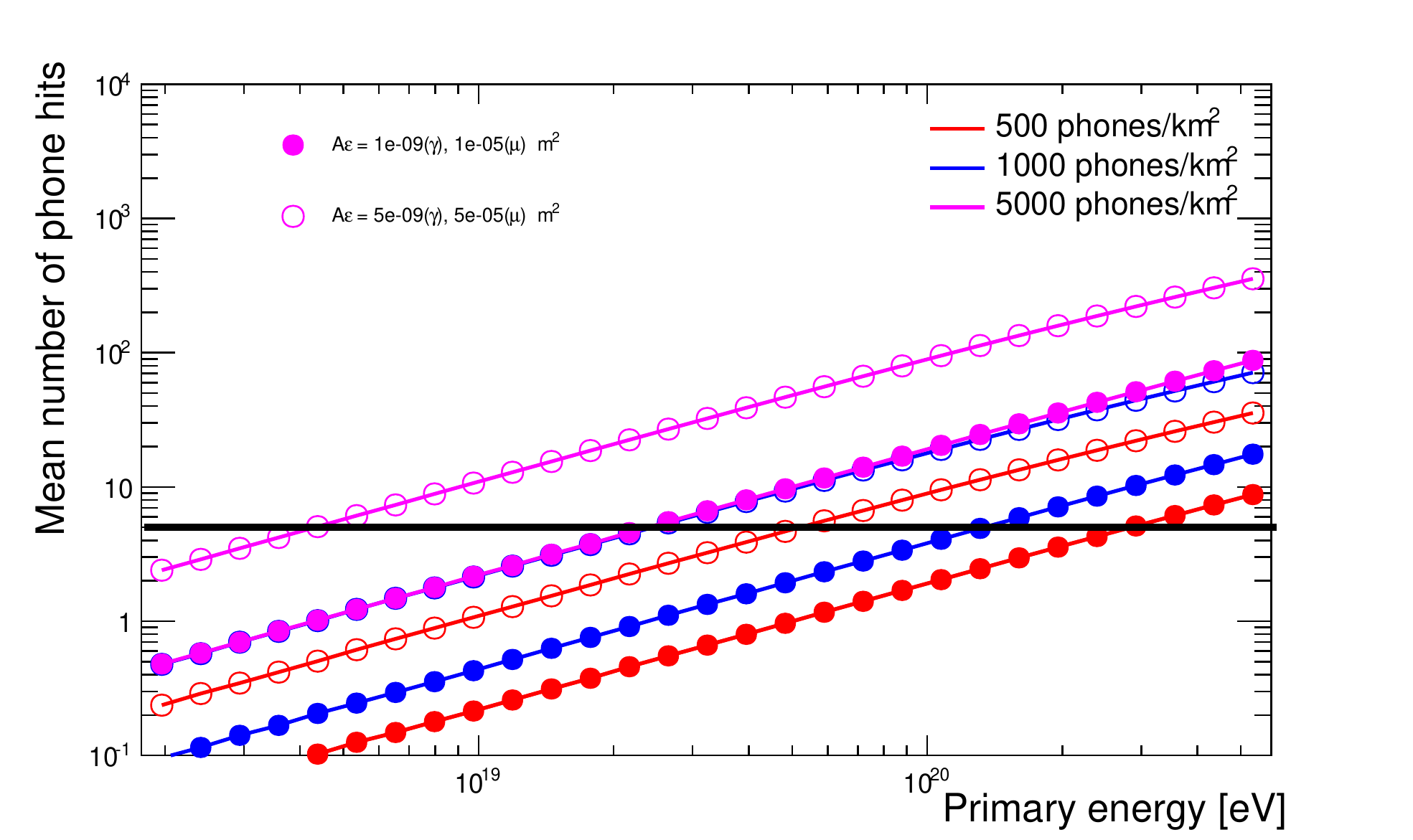}
\includegraphics[width=3in]{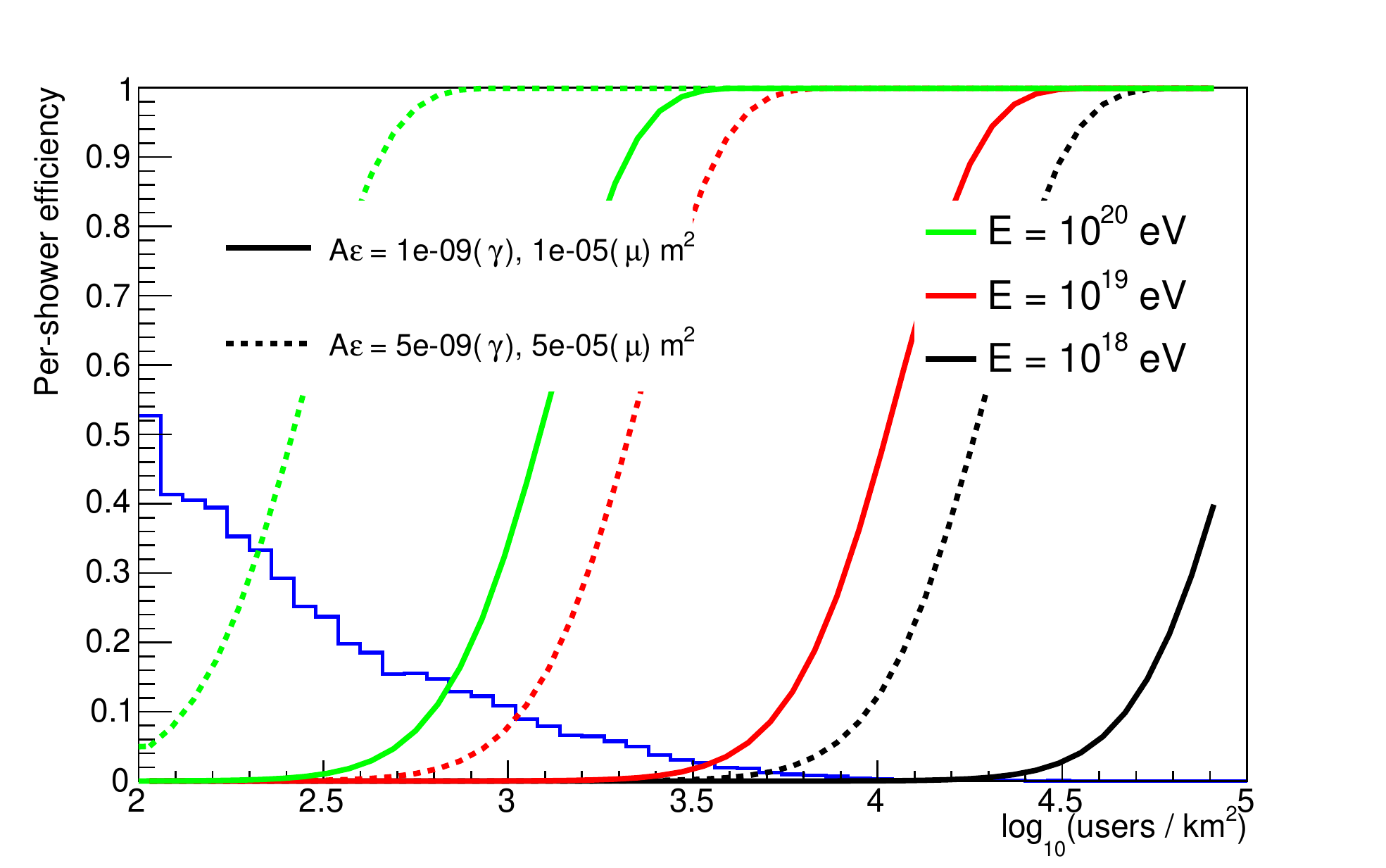}
\caption{ Top, mean number of phones registering particle hits per shower versus incident primary particle energy,  for two choices of per-phone area times efficiency ($A\epsilon$)  and three examples of user adoption density. Bottom, per-shower efficiency as a function of the density of users in a 1-km$^{2}$ region; overlaid is the relative distribution of actual population densities~\cite{grumpy} with arbitrary normalization.}
\label{fig:num}
\end{figure}

The per-shower efficiency is calculated in simulation by sampling
randomly placed phones in the path of a shower and determining the
number of phones which register a hit. To suppress the background from uncorrelated hits, we choose a benchmark requirement that at least five phones
register a hit to be considered a shower candidate; see
Fig.~\ref{fig:num}.  The per-shower efficiency is then the fraction of
showers which have at least five phones registering hits.  The
efficiency is a strong function of the density of participating devices, see bottom
panel of Fig~\ref{fig:num}. In addition, the per-shower efficiency
rises with incident particle energy due to the increase in the number
of particles in the shower.  The dominant contribution at all angles
is from muons, due to their much larger efficiency, despite their
rarity with respect to photons in vertical showers.

\begin{figure}[hbt!]
\includegraphics[width=3in]{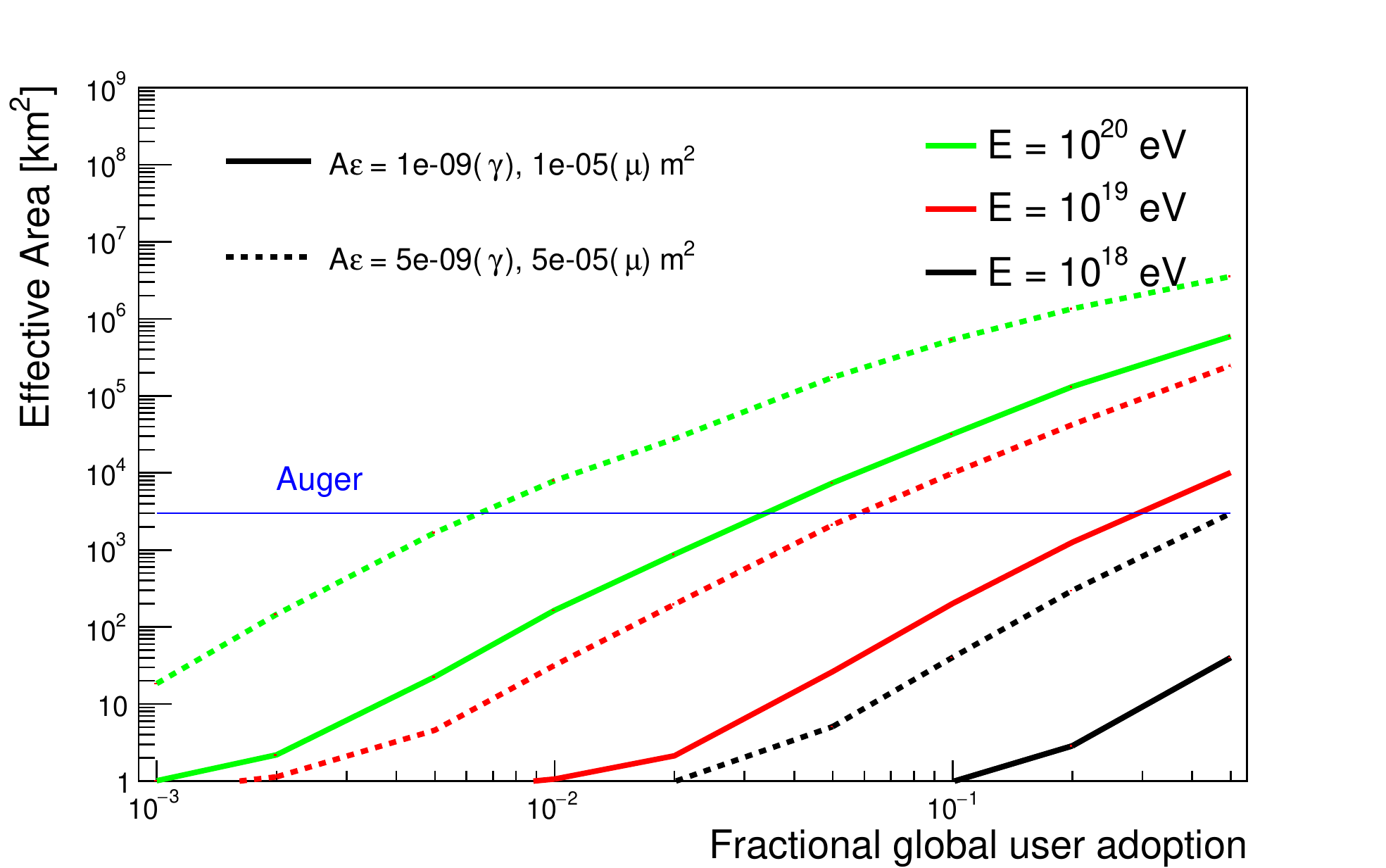}
\includegraphics[width=3in]{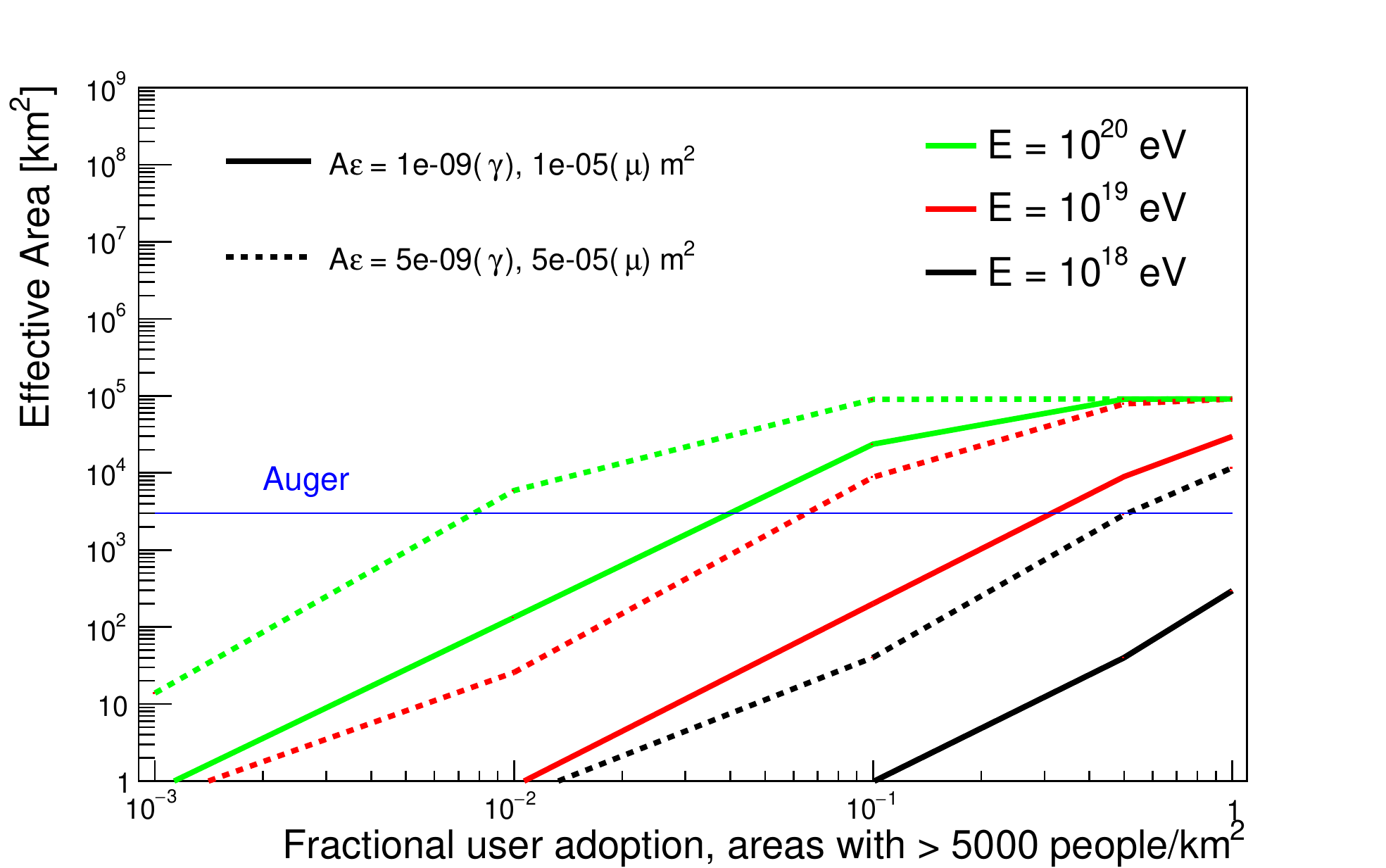}
\caption{ Top, effective area as a function of the fraction of the global population which participates, for several choices of primary cosmic-ray energy and two choices of per-phone efficiency benchmarks. Bottom, effective area as a function of the fraction of the population in very high density regions (areas with $>5000$ people per km$^{2}$).}
\label{fig:perf}
\end{figure}

\begin{figure}[hbt!]
\includegraphics[width=3in]{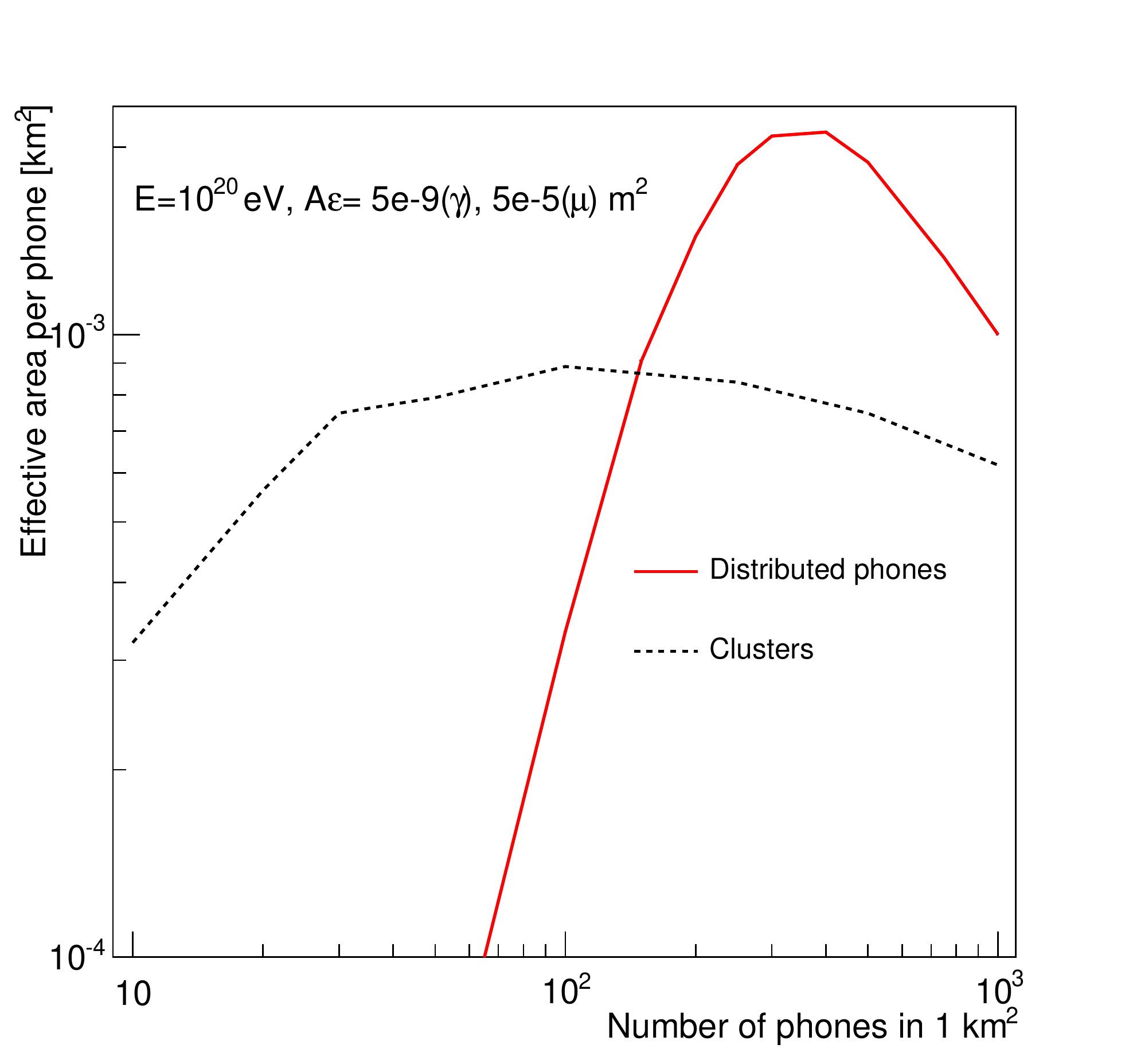}
\caption{ Effective area per phone for showers with $E=10^{20}$ eV under the optimistic $A\epsilon$ scenario, as a function of the number of phones in a 1-km$^{2}$ square under two clustering possibilities: either the phones are randomly distributed, or grouped in a small cluster. For high densities ($N>100$/km$^{2}$), distributed phones have the largest contributions. For low densities ($N<100$/km$^{2}$), clusters can make significant contributions.}
\label{fig:clus}
\end{figure}

The observing power of state-of-the-art dedicated ground arrays is determined by the {\it effective area}, which is the integral of the per-shower efficiency over the detector area.  We leave considerations of the angular field of view for future studies. For Auger, efficiency for showers above $E=10^{20}$ eV is effectively 100\%, giving an effective area of 3000 km$^{2}$, the size of the instrumented area. To achieve a similar efficiency and effective area would require a density of approximately 400 dedicated devices per km$^{2}$ for the optimistic $A\epsilon$ scenario, distributed over 3000 km$^{2}$, for a total of 1.2M devices. 

In practice, we expect the location of devices to be tied to existing population distributions.  To assess the size of a  network needed to match the effective area of Auger, we perform an analysis similar to Ref.~\cite{Unger:2015wka} and 
calculate the effective area of a network composed of the devices of a fraction $f$ of the world population.  Population data~\cite{grumpy} at the granularity of 30 arc-sec$^{2}$ (0.5-0.8 km$^{2}$ depending on latitude) are overlaid with the efficiency curves in the top pane of Fig~\ref{fig:num}.  In Fig.~\ref{fig:perf}, the effective area for several cases are shown as a function of the fractional user adoption. In the optimistic $A\epsilon$ scenario, less than 1\% user adoption ($\approx$ 45M users) is needed to match the effective area of Auger for UHECRs.\footnote{ A similar calculation~\cite{Unger:2015wka} yielded a much smaller estimate of the effective area at $E=10^{20}$ eV due to an assumption of zero per-shower efficiency for device densities less than $1000$/km$^{2}$; see Ref.~\cite{Unger:2015wka} for discussion of this assumption. This smaller estimate of the effective area leads to a corresponding larger estimate of the required user fraction.} This effective area is dominated by high-density areas due to their higher per-shower efficiency; therefore, the deployment strategy should focus on recruiting contributors in high-density areas. 
The bottom pane of Fig.~\ref{fig:perf} shows that the same effective area can be achieved with the same user fraction when restricted to areas with population density greater than $5000/$km$^{2}$, which corresponds to $\approx$ 7M users.   The effective area is only significantly reduced by restricting to these high-density areas when the fractional user adoption is high.

In low population density areas, small clusters of phones can also contribute significantly to the total effective area, though with degraded energy and angular resolution.  
The per-shower efficiency is a strongly non-linear function of the number and arrangement of phones due to the requirement of a 5-phone coincidence.
Figure~\ref{fig:clus} shows that clusters of 30 or more phones provide an effective area per phone that is within a factor of 2-3 of a uniformly random distribution with 400 phones per km${^2}$. For example, 134k (15k)  localized clusters of 30 (250)  phones can also achieve an effective area of 3000 km$^{2}$.
Such clusters might be constructed from donated phones no longer in active use, which motivates a deployment strategy that includes partnerships with schools and science clubs around the world.

Note that these calculations assume continuous operation; some degradation of observational power is expected, as some phones may only join the network during night-time charging.  On the other hand, users may dedicate devices no longer in active use.  The observational power of such a network clearly hinges on the level of user adoption and continued participation.

\section*{Conclusions}

We propose a novel strategy for observing air showers due to ultra-high energy cosmic rays: an array composed of smartphones running a dedicated app.  We have measured the per-phone sensitivity to the particles which comprise the showers and estimated the number of phones needed to achieve observing power to rival the most sensitive current observatories.

Building an installed user base of more than 1M devices operating reliably poses a social and organizational challenge. We have begun to address these by reducing the barriers to participation via unobtrusive operation, and providing incentives for users.

A large network of devices would have unprecedented observing power at energies above 10$^{20}$ eV, where current ground arrays become saturated~\cite{augercurve}. Lack of observations of UHECRs above this energy could therefore provide powerful limits on the incident flux.

Such a world-wide network of devices sensitive to muons and photons could also have many other potential uses, such as monitoring local radiation levels. In addition, this global network would be the first of its kind, opening a new observational window to unanticipated processes.

\section*{Acknowledgements}

We thank Davide Gerbaudo, Tatiana Rodgriguez, Josh Cogliati, Rocky Dendo, Steve Barwick, Gourang Yodh, John Beacom, Jonathan Wallick for useful conversations, alpha testing and source calibration.

\bibliographystyle{apsrev}
\bibliography{paper}

\begin{thebibliography}{28}
\expandafter\ifx\csname natexlab\endcsname\relax\def\natexlab#1{#1}\fi
\expandafter\ifx\csname bibnamefont\endcsname\relax
  \def\bibnamefont#1{#1}\fi
\expandafter\ifx\csname bibfnamefont\endcsname\relax
  \def\bibfnamefont#1{#1}\fi
\expandafter\ifx\csname citenamefont\endcsname\relax
  \def\citenamefont#1{#1}\fi
\expandafter\ifx\csname url\endcsname\relax
  \def\url#1{\texttt{#1}}\fi
\expandafter\ifx\csname urlprefix\endcsname\relax\def\urlprefix{URL }\fi
\providecommand{\bibinfo}[2]{#2}
\providecommand{\eprint}[2][]{\url{#2}}

\bibitem[{\citenamefont{Abraham et~al.}(2008{\natexlab{a}})}]{Abraham:2007si}
\bibinfo{author}{\bibfnamefont{J.}~\bibnamefont{Abraham}} \bibnamefont{et~al.}
  (\bibinfo{collaboration}{Pierre Auger Collaboration}),
  \bibinfo{journal}{Astropart.Phys.} \textbf{\bibinfo{volume}{29}},
  \bibinfo{pages}{188} (\bibinfo{year}{2008}{\natexlab{a}}),
  \eprint{0712.2843}.

\bibitem[{\citenamefont{Abraham et~al.}(2007)}]{Abraham:2007bb}
\bibinfo{author}{\bibfnamefont{J.}~\bibnamefont{Abraham}} \bibnamefont{et~al.}
  (\bibinfo{collaboration}{Pierre Auger Collaboration}),
  \bibinfo{journal}{Science} \textbf{\bibinfo{volume}{318}},
  \bibinfo{pages}{938} (\bibinfo{year}{2007}), \eprint{0711.2256}.

\bibitem[{\citenamefont{Bell}(1978)}]{Bell:1978zc}
\bibinfo{author}{\bibfnamefont{A.~R.} \bibnamefont{Bell}},
  \bibinfo{journal}{Mon.Not.Roy.Astron.Soc.} \textbf{\bibinfo{volume}{182}},
  \bibinfo{pages}{147} (\bibinfo{year}{1978}).

\bibitem[{\citenamefont{Blandford and Eichler}(1987)}]{Blandford:1987pw}
\bibinfo{author}{\bibfnamefont{R.}~\bibnamefont{Blandford}} \bibnamefont{and}
  \bibinfo{author}{\bibfnamefont{D.}~\bibnamefont{Eichler}},
  \bibinfo{journal}{Phys.Rept.} \textbf{\bibinfo{volume}{154}},
  \bibinfo{pages}{1} (\bibinfo{year}{1987}).

\bibitem[{\citenamefont{Waxman}(1995)}]{Waxman:1995vg}
\bibinfo{author}{\bibfnamefont{E.}~\bibnamefont{Waxman}},
  \bibinfo{journal}{Phys.Rev.Lett.} \textbf{\bibinfo{volume}{75}},
  \bibinfo{pages}{386} (\bibinfo{year}{1995}), \eprint{astro-ph/9505082}.

\bibitem[{\citenamefont{Weiler}(1999)}]{Weiler:1997sh}
\bibinfo{author}{\bibfnamefont{T.~J.} \bibnamefont{Weiler}},
  \bibinfo{journal}{Astropart.Phys.} \textbf{\bibinfo{volume}{11}},
  \bibinfo{pages}{303} (\bibinfo{year}{1999}), \eprint{hep-ph/9710431}.

\bibitem[{\citenamefont{Abbasi et~al.}(2004)}]{Abbasi:2002ta}
\bibinfo{author}{\bibfnamefont{R.}~\bibnamefont{Abbasi}} \bibnamefont{et~al.}
  (\bibinfo{collaboration}{High Resolution Fly's Eye Collaboration}),
  \bibinfo{journal}{Phys.Rev.Lett.} \textbf{\bibinfo{volume}{92}},
  \bibinfo{pages}{151101} (\bibinfo{year}{2004}), \eprint{astro-ph/0208243}.

\bibitem[{\citenamefont{Takeda et~al.}(1998)\citenamefont{Takeda, Hayashida,
  Honda, Inoue, Kadota et~al.}}]{Takeda:1998ps}
\bibinfo{author}{\bibfnamefont{M.}~\bibnamefont{Takeda}},
  \bibinfo{author}{\bibfnamefont{N.}~\bibnamefont{Hayashida}},
  \bibinfo{author}{\bibfnamefont{K.}~\bibnamefont{Honda}},
  \bibinfo{author}{\bibfnamefont{N.}~\bibnamefont{Inoue}},
  \bibinfo{author}{\bibfnamefont{K.}~\bibnamefont{Kadota}},
  \bibnamefont{et~al.}, \bibinfo{journal}{Phys.Rev.Lett.}
  \textbf{\bibinfo{volume}{81}}, \bibinfo{pages}{1163} (\bibinfo{year}{1998}),
  \eprint{astro-ph/9807193}.

\bibitem[{\citenamefont{Abraham et~al.}(2008{\natexlab{b}})}]{Abraham:2008ru}
\bibinfo{author}{\bibfnamefont{J.}~\bibnamefont{Abraham}} \bibnamefont{et~al.}
  (\bibinfo{collaboration}{Pierre Auger Collaboration}),
  \bibinfo{journal}{Phys.Rev.Lett.} \textbf{\bibinfo{volume}{101}},
  \bibinfo{pages}{061101} (\bibinfo{year}{2008}{\natexlab{b}}),
  \eprint{0806.4302}.

\bibitem[{\citenamefont{Greisen}(1966)}]{Greisen:1966jv}
\bibinfo{author}{\bibfnamefont{K.}~\bibnamefont{Greisen}},
  \bibinfo{journal}{Phys.Rev.Lett.} \textbf{\bibinfo{volume}{16}},
  \bibinfo{pages}{748} (\bibinfo{year}{1966}).

\bibitem[{\citenamefont{Zatsepin and Kuzmin}(1966)}]{Zatsepin:1966jv}
\bibinfo{author}{\bibfnamefont{G.}~\bibnamefont{Zatsepin}} \bibnamefont{and}
  \bibinfo{author}{\bibfnamefont{V.}~\bibnamefont{Kuzmin}},
  \bibinfo{journal}{JETP Lett.} \textbf{\bibinfo{volume}{4}},
  \bibinfo{pages}{78} (\bibinfo{year}{1966}).

\bibitem[{\citenamefont{Aab et~al.}(2014)}]{Aab:2014ila}
\bibinfo{author}{\bibfnamefont{A.}~\bibnamefont{Aab}} \bibnamefont{et~al.}
  (\bibinfo{collaboration}{Telescope Array Collaboration, Pierre Auger
  Collaboration}), \bibinfo{journal}{Astrophys.J.}  (\bibinfo{year}{2014}),
  \eprint{1409.3128}.

\bibitem[{\citenamefont{Abu-Zayyad et~al.}(2013)}]{AbuZayyad:2012ru}
\bibinfo{author}{\bibfnamefont{T.}~\bibnamefont{Abu-Zayyad}}
  \bibnamefont{et~al.} (\bibinfo{collaboration}{Telescope Array}),
  \bibinfo{journal}{Astrophys.J.} \textbf{\bibinfo{volume}{768}},
  \bibinfo{pages}{L1} (\bibinfo{year}{2013}), \eprint{1205.5067}.

\bibitem[{\citenamefont{Hayashida et~al.}(1999)}]{Hayashida:1998qb}
\bibinfo{author}{\bibfnamefont{N.}~\bibnamefont{Hayashida}}
  \bibnamefont{et~al.} (\bibinfo{collaboration}{AGASA Collaboration}),
  \bibinfo{journal}{Astropart.Phys.} \textbf{\bibinfo{volume}{10}},
  \bibinfo{pages}{303} (\bibinfo{year}{1999}), \eprint{astro-ph/9807045}.

\bibitem[{erg()}]{ergo}
\urlprefix\url{http://www.ergotelescope.org/}.

\bibitem[{\citenamefont{Cogliati et~al.}(2014)\citenamefont{Cogliati, Derr, and
  Wharton}}]{Cogliati:2014uua}
\bibinfo{author}{\bibfnamefont{J.~J.} \bibnamefont{Cogliati}},
  \bibinfo{author}{\bibfnamefont{K.~W.} \bibnamefont{Derr}}, \bibnamefont{and}
  \bibinfo{author}{\bibfnamefont{J.}~\bibnamefont{Wharton}}
  (\bibinfo{year}{2014}), \eprint{1401.0766}.

\bibitem[{\citenamefont{{Smith} et~al.}(2002)\citenamefont{{Smith}, {McDonald},
  {Hurley}, {Holland}, {Groom}, {Brown}, {Gilmore}, {Stover}, and
  {Wei}}}]{2002SPIE.4669..172S}
\bibinfo{author}{\bibfnamefont{A.~R.} \bibnamefont{{Smith}}},
  \bibinfo{author}{\bibfnamefont{R.~J.} \bibnamefont{{McDonald}}},
  \bibinfo{author}{\bibfnamefont{D.~C.} \bibnamefont{{Hurley}}},
  \bibinfo{author}{\bibfnamefont{S.~E.} \bibnamefont{{Holland}}},
  \bibinfo{author}{\bibfnamefont{D.~E.} \bibnamefont{{Groom}}},
  \bibinfo{author}{\bibfnamefont{W.~E.} \bibnamefont{{Brown}}},
  \bibinfo{author}{\bibfnamefont{D.~K.} \bibnamefont{{Gilmore}}},
  \bibinfo{author}{\bibfnamefont{R.~J.} \bibnamefont{{Stover}}},
  \bibnamefont{and} \bibinfo{author}{\bibfnamefont{M.}~\bibnamefont{{Wei}}}, in
  \emph{\bibinfo{booktitle}{Sensors and Camera Systems for Scientific,
  Industrial, and Digital Photography Applications III}}, edited by
  \bibinfo{editor}{\bibfnamefont{M.~M.} \bibnamefont{{Blouke}}},
  \bibinfo{editor}{\bibfnamefont{J.}~\bibnamefont{{Canosa}}}, \bibnamefont{and}
  \bibinfo{editor}{\bibfnamefont{N.}~\bibnamefont{{Sampat}}}
  (\bibinfo{year}{2002}), vol. \bibinfo{volume}{4669} of
  \emph{\bibinfo{series}{Society of Photo-Optical Instrumentation Engineers
  (SPIE) Conference Series}}, pp. \bibinfo{pages}{172--183}.

\bibitem[{\citenamefont{DECO}(2012)}]{decoaps}
\bibinfo{author}{\bibnamefont{DECO}}, \bibinfo{journal}{APS News}
  \textbf{\bibinfo{volume}{21}}, \bibinfo{pages}{3} (\bibinfo{year}{2012}).

\bibitem[{web()}]{website}
\urlprefix\url{http://crayfis.io/}.

\bibitem[{\citenamefont{Heck et~al.}(1998)\citenamefont{Heck, Schatz, Thouw,
  Knapp, and Capdevielle}}]{Heck:1998vt}
\bibinfo{author}{\bibfnamefont{D.}~\bibnamefont{Heck}},
  \bibinfo{author}{\bibfnamefont{G.}~\bibnamefont{Schatz}},
  \bibinfo{author}{\bibfnamefont{T.}~\bibnamefont{Thouw}},
  \bibinfo{author}{\bibfnamefont{J.}~\bibnamefont{Knapp}}, \bibnamefont{and}
  \bibinfo{author}{\bibfnamefont{J.}~\bibnamefont{Capdevielle}}
  (\bibinfo{year}{1998}).

\bibitem[{\citenamefont{Ostapchenko}(2011)}]{Ostapchenko:2010vb}
\bibinfo{author}{\bibfnamefont{S.}~\bibnamefont{Ostapchenko}},
  \bibinfo{journal}{Phys.Rev.} \textbf{\bibinfo{volume}{D83}},
  \bibinfo{pages}{014018} (\bibinfo{year}{2011}), \eprint{1010.1869}.

\bibitem[{\citenamefont{Matis et~al.}(2003)\citenamefont{Matis, Bieser, Rai,
  Retiere, Ritter et~al.}}]{Matis:2002jv}
\bibinfo{author}{\bibfnamefont{H.~S.} \bibnamefont{Matis}},
  \bibinfo{author}{\bibfnamefont{F.}~\bibnamefont{Bieser}},
  \bibinfo{author}{\bibfnamefont{G.}~\bibnamefont{Rai}},
  \bibinfo{author}{\bibfnamefont{F.}~\bibnamefont{Retiere}},
  \bibinfo{author}{\bibfnamefont{H.~G.} \bibnamefont{Ritter}},
  \bibnamefont{et~al.}, \bibinfo{journal}{IEEE Trans.Nucl.Sci.}
  \textbf{\bibinfo{volume}{50}}, \bibinfo{pages}{1020} (\bibinfo{year}{2003}),
  \eprint{nucl-ex/0212019}.

\bibitem[{\citenamefont{Kleinfelder et~al.}(2001)\citenamefont{Kleinfelder,
  Lim, Lui, and El~Gamal}}]{Kleinf2002}
\bibinfo{author}{\bibfnamefont{S.}~\bibnamefont{Kleinfelder}},
  \bibinfo{author}{\bibfnamefont{S.-K.} \bibnamefont{Lim}},
  \bibinfo{author}{\bibfnamefont{X.}~\bibnamefont{Lui}}, \bibnamefont{and}
  \bibinfo{author}{\bibfnamefont{A.}~\bibnamefont{El~Gamal}},
  \bibinfo{journal}{IEEE Journal of Solid-State Circuits}
  \textbf{\bibinfo{volume}{36}} (\bibinfo{year}{2001}).

\bibitem[{\citenamefont{Agostinelli et~al.}(2003)}]{Agostinelli:2002hh}
\bibinfo{author}{\bibfnamefont{S.}~\bibnamefont{Agostinelli}}
  \bibnamefont{et~al.} (\bibinfo{collaboration}{GEANT4}),
  \bibinfo{journal}{Nucl.Instrum.Meth.} \textbf{\bibinfo{volume}{A506}},
  \bibinfo{pages}{250} (\bibinfo{year}{2003}).

\bibitem[{\citenamefont{Grieder}(2010)}]{grieder}
\bibinfo{author}{\bibfnamefont{P.}~\bibnamefont{Grieder}},
  \emph{\bibinfo{title}{Extensive Air Showers}} (\bibinfo{publisher}{Springer},
  \bibinfo{year}{2010}).

\bibitem[{gru()}]{grumpy}
\urlprefix\url{http://sedac.ciesin.columbia.edu/data/set/grump-v1-population-density}.

\bibitem[{\citenamefont{Unger and Farrar}(2015)}]{Unger:2015wka}
\bibinfo{author}{\bibfnamefont{M.}~\bibnamefont{Unger}} \bibnamefont{and}
  \bibinfo{author}{\bibfnamefont{G.}~\bibnamefont{Farrar}}
  (\bibinfo{year}{2015}), \eprint{1505.04777}.

\bibitem[{\citenamefont{Aab et~al.}(2013)}]{augercurve}
\bibinfo{author}{\bibfnamefont{A.}~\bibnamefont{Aab}} \bibnamefont{et~al.}
  (\bibinfo{collaboration}{Pierre Auger Collaboration}) (\bibinfo{year}{2013}),
  \eprint{1307.5059}.

\end{thebibliography}

\end{document}